\documentclass[twocolumn]{aastex631}
\usepackage{amsmath}
\usepackage{multirow}

\usepackage{CJK}         
\usepackage{comment}
\usepackage{array}
\usepackage{booktabs}

\newcommand{\pfrac}[2]{\left( \frac{#1}{#2} \right)}

\newcommand{\rt}{r_{\rm t}}
\newcommand{\MBH}{M_{\rm BH}}

\begin{document}
\begin{CJK*}{UTF8}{gbsn}

\title{Star-Disk Collisions II: Debris Stream Dynamics\\ and Implications for QPEs and Other Transients Near SMBHs}

\author[0000-0003-3024-7218]{Philippe Z. Yao}
\affil{Department of Astrophysical Sciences, Princeton University, Peyton Hall, Princeton, NJ 08544, USA}

\author[0000-0001-9185-5044]{Eliot Quataert}
\affil{Department of Astrophysical Sciences, Princeton University, Peyton Hall, Princeton, NJ 08544, USA}

\author[0000-0002-2624-3399]{Yan-Fei Jiang (姜燕飞)}
\affil{Center for Computational Astrophysics, Flatiron Institute, New York, NY 10010, USA}

\author[0000-0002-8304-1988]{Itai Linial}\thanks{NASA Einstein Fellow}
\affil{Department of Physics and Columbia Astrophysics Laboratory, Columbia University, New York, NY 10027, USA}
\affiliation{Center for Cosmology and Particle Physics, Physics Department, New York University, New York, NY 10003, USA}

\begin{abstract}
Quasi-periodic eruptions (QPEs) are repeating soft X-ray nuclear transients with recurrence times of hours--days and flare duty cycles of $\sim$10--20\%. Many aspects of QPEs can be modeled as a stellar-mass orbiter that intersects an accretion disk producing a shocked debris cloud and a flare of radiation. We present three-dimensional \texttt{Athena++} hydrodynamic simulations of star-disk interactions around a $10^{6}\,M_\odot$ supermassive black hole, including the black hole's tidal potential, the disk's Keplerian rotation, and orbital periods similar to those observed. After each disk encounter, freshly stripped stellar debris exits the Hill sphere to form an extended, asymmetric, roughly triaxial stream. Subsequent stream-disk collisions shock both stellar debris and disk gas to high specific energies and drive a wind-like outflow. At larger orbital periods the shocked stellar debris dominates the high specific energy debris, while at shorter orbital periods the shocked disk energy can be similar. From the shocked stellar mass measured in the simulations over time, we infer flare durations set by the time it takes the stellar debris stream to collide with the disk, consistent with the observed constant duty cycle of $\sim$10--20\%, independent of orbital period.  The total shocked debris energy is consistent with QPE flare energetics. Our results favor one observable flare per stellar orbit except perhaps at the shortest orbital periods where the shocked star and disk energetics can be similar. Variations in the stream's center of mass relative to the star, the stream density, and other properties can produce diverse changes in the time of the flare's peak relative to the time of the star-disk collision. We discuss the implications of our results for QPE timing and for other transients in galactic nuclei.
\end{abstract}


\section{Introduction} \label{sec:intro}

Quasi-periodic eruptions (QPEs) are repeating nuclear transients near supermassive black holes (SMBHs). They recur every few hours to days, with each flare lasting on average about $10-20\%$ of their recurrence time. QPEs are found near lower-mass SMBHs between $\sim 10^5-10^7 M_\odot$ \citep{Wevers2022}. They are almost only observed in soft X-rays\footnote{The remarkable source Ansky is currently unique in having UV variability that correlates with its X-ray flares \citep{Guo2026}.} with blackbody temperatures of $kT\sim 50-200 \rm eV$ and bolometric luminosities of $L_{\rm bol}\sim 10^{41}-10^{44} \rm erg/s$, an order of magnitude or more above a cooler quiescent level. Roughly half of the confirmed QPE sources have been uncovered through blind X-ray searches \citep{Miniutti2019, Giustini2020, Chakraborty2021, Arcodia2021, Arcodia2024, Arcodia2025}. The remainder were revealed by archival mining and by targeted X-ray follow-up of longer-duration nuclear transients, often tidal disruption events (TDEs) \citep{Quintin2023, Bykov2025, Nicholl2024, HernandezGarcia2025, Chakraborty2025, Baldini2026}. 

QPE sources show diversity in recurrence time, burst morphology, and long-term evolution, hinting that the underlying engine may be sensitive to the properties of the SMBH and its near environment. While a broad range of theoretical scenarios have been proposed, including disk instability models (e.g., \citealt{Raj2021,Kaur2023}), we focus here on models in which a star on a mildly eccentric orbit produces flares via its interactions with a gaseous accretion disk once or twice per orbit (see, e.g., \citealt{Xian2021, Sukova2021,Krolik2022,Zhao2022, Linial2023,Linial2024b,Lu2023,Tagawa2023, Franchini2023}). The observed diversity in sources can then be understood in terms of black-hole mass and spin, disk structure, orbital eccentricity, and the nature of the interacting secondary. The star plausibly originates from an extreme mass-ratio inspiral (EMRI) produced by Hills breakup of a tight binary on an initially highly eccentric orbit, later circularized by gravitational waves (GWs) and disk drag \citep[e.g.,][]{Linial2023a}. The source of the disk is speculated to originate from TDEs occurring every $10^4-10^5$ years \citep[e.g.][]{Stone2016, Yao2023}, which spreads to enclose EMRI orbits. This connection between TDEs and QPEs was initially suspected in observations of GSN 069 \citep{Miniutti2019}, where the quiescent luminosity of the source resembles a very slowly decaying TDE, and was later formalized in \citet{Linial2023} where they show that a QPE-TDE connection is a natural consequence of EMRI and TDE formation rates. One of the strongest pieces of evidence for this connection comes from the discovery of QPEs 4 years after the optically selected, spectroscopically confirmed TDE AT 2019qiz \citep{Nicholl2024}. 

Understanding the origin of QPE flares requires careful numerical work to untangle the detailed hydrodynamic and emission processes in these complex systems. Motivated by the star-disk collision interpretation, \citet{Vurm2025} develop spherically symmetric Monte Carlo radiation-transport simulations that explicitly track radiation-mediated shock formation, photon production, Comptonization, and photon escape from the expanding shocked debris, demonstrating that this framework can yield QPE-like soft X-ray eruptions with characteristic spectral evolution.  In parallel, \citet{Huang2025} carry out 2D \texttt{Athena++} radiation-hydrodynamic simulations (both gray and multigroup transfer from \citealt{Jiang2022}) of a gravitating polytropic star impacting an optically thick disk column in the star's comoving frame, enabling a more self-consistent connection between the collision hydrodynamics, opacity/reprocessing, and the resulting time-dependent SED. 

The radiative models are, however, sensitive to the uncertain hydrodynamics producing the observed flares, which is the focus of this paper: we study the tidal evolution of the stripped debris under the SMBH's gravity and quantify the energetics of the subsequent collisions between the rotating accretion
disk and the elongated debris stream.

\citet{Yao2025} investigated the hydrodynamic effects on the star of repeated star-disk encounters. A key outcome is that repeated impacts qualitatively change the effective ``collider”: shock heating inflates the bound gas envelope around the star to $\sim(2$--$3)\,R_\odot$ and the surrounding unbound debris cloud further enlarges the cross section for subsequent interactions. Our initial calculations in \citet{Yao2025} did not include the black hole's tidal gravity, but we argued that its influence would lead to a tidally stretched debris stream that would dominate the radiation in star-disk QPE models. \citet{Linial2025} then developed this into an analytic model of QPE flares. 

The goal of this paper is to build on the debris-disk collision model proposed in \citet{Yao2025} and explored analytically in detail in \citet{Linial2025}. The debris-disk collision model presents opportunities to address several observational challenges facing the pure star-disk collision models \citep{Guo2026a,Mummery2025}. Our primary focus is on placing quantitative constraints on three aspects of the collision dynamics that bear on these observational challenges: (1) the energy and mass of shocked debris compared to those of the shocked disk, (2) the duration for which the debris is shocked and its connection to the flare duration, and (3) the dependence on orbital period at fixed black hole mass. 

The remainder of this paper is organized as follows. In \S \ref{sec:sim} we describe our numerical setup. \S \ref{sec:results} presents the main simulation results, focusing on the mass loss, debris stream formation, energetics of shocked material, and flare signatures. We conclude in \S \ref{sec:discussion} with a summary of our principal findings, a discussion of caveats, and the implications of this debris-disk interaction model to the broader observational signatures of QPEs.

\section{Simulation Methods} \label{sec:sim}

\begin{figure*}
    \centering
    \includegraphics[width=\textwidth]{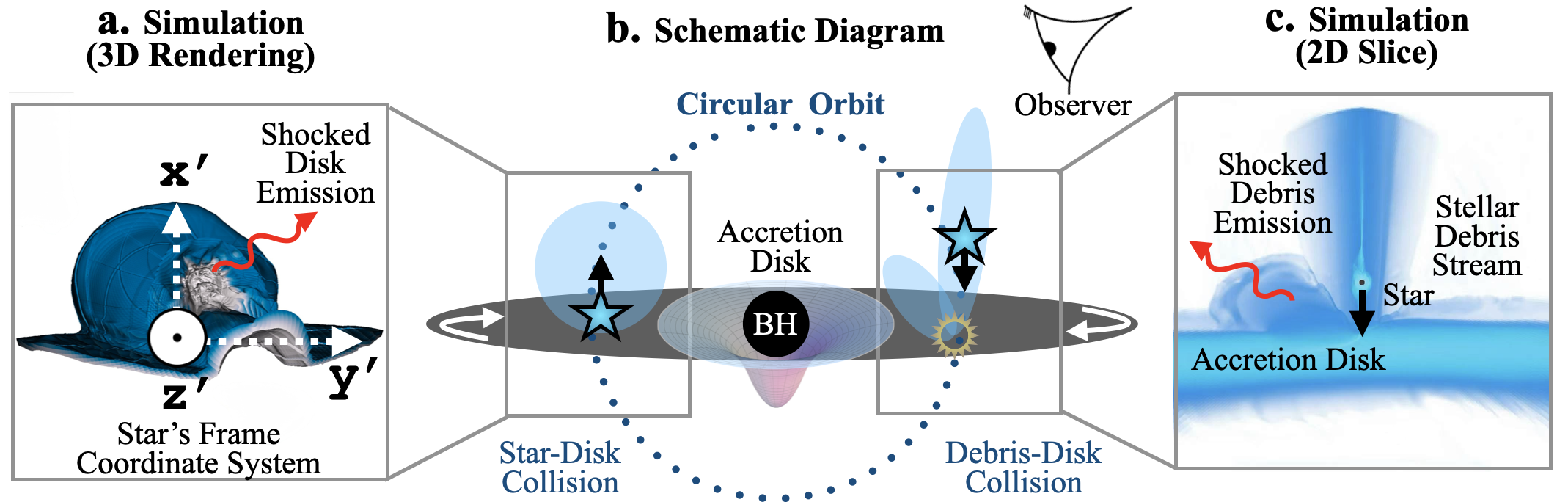} 
    \caption{\textbf{(a)} 3D simulation rendering of shocked disk produced in a star-debris-disk collision, also showing the coordinate system referenced in \S \ref{sec:sim} where $y'$ points in the direction of the black hole and the disk moves toward the star along the $-x'$ direction \textbf{(b)} Schematic illustration of the main physical ingredients in our simulation setup for modeling QPEs as a stellar companion to an SMBH repeatedly colliding with a rotating accretion disk. Only processes on the side of the observer are illustrated for clarity. \textbf{(c)} 2D simulation snapshot of shocked stellar debris, where a wind-like outflow is launched as elongated stellar debris is shocked by the collision with the accretion disk. The orientation of the wind-like outflow is tilted from and almost perpendicular to the orbital plane due to shocked material advected by the rotation of the disk.}
    \label{fig:schematic}
\end{figure*}

We carried out 3D simulations using the \texttt{Athena++} ideal hydrodynamics code \citep{Stone2020} in spherical polar coordinates with a statically refined logarithmic grid. Compared to our previous set of 2D simulations presented in \citet{Yao2025}, in addition to the extra dimension, we include the following additional physics, illustrated in Figure \ref{fig:schematic}:
\begin{itemize}
    \item The tidal potential of a $10^6 M_{\odot}$ SMBH, including the centrifugal and Coriolis forces in the co-rotating frame of the star and the BH. This allows us to study the full orbital evolution of stripped stellar debris, as a function of the star's orbital period.

    \item The Keplerian rotation of the accretion disk, and its tilt relative to the stellar orbital plane. This affects both the magnitude and direction of the star-disk collision. 

    \item A realistic orbital period as a function of the orbital distance to the SMBH. This allows us to compare the diverse timing behavior of QPEs at different recurrence times with star-disk collisions taking place across a range of orbital periods. 
\end{itemize}

The initial profile of the star, positioned at the geometric center of our spherical grid, is not an unperturbed solar model, but a solar-like $
\gamma = 5/3$ polytrope that has experienced multiple star-disk collisions (see \S 2 of \citealt{Yao2025} for a more detailed discussion). Specifically, we take the spherically averaged density and pressure profile of the bound star from the fiducial set of 2D simulations after 5 collisions. These collisions change the surface layers of the star drastically by inflating them via shock heating. With this initial condition, we significantly reduce the number of collisions we need for convergence in the much more expensive 3D simulations. 

Since only a thin shell of the stellar envelope ($r > 0.9 R_{\odot}$) is present in our simulation domain, we use a fixed point-mass potential at the center of the grid for the star's gravity. For each orbital separation, we perform 3 collisions until the mass loss per collision stabilizes. Every half an orbit, the disk is created at $r_{\rm disk}\sim100R_{\odot}$ away from the star to capture as much of the stream as possible. Passive scalars that evolve with the fluid but do not alter the fluid's behavior are used as tracer dyes to distinguish between star and disk material.  Stellar debris outside $r_{\rm disk}$ is artificially removed. The disk has an initial tilt $\theta_{\rm disk}$ in the frame of the star co-rotating with the SMBH. In our logarithmically spaced grid, this leads to a compromise that the disk is initialized with a lower resolution, which can numerically diffuse before colliding the star. As a result, we opt for a thicker ($\sim4\times$) accretion disk with $H_{\rm disk}\sim5.6R_{\odot}$ than in \citet{Yao2025} to minimize this effect. Numerical diffusion did not pose a significant challenge in \citet{Yao2025} both because the disk is introduced much closer to the star, and because higher resolution simulations are much less computationally expensive in 2D. Our convergence test shows that for properties of the shocked stellar debris, this thicker disk in our 3D simulations yields negligible difference.   

\subsection{Black Hole's Tidal Potential}

In our setup, the SMBH gravity and the effects of rotation are incorporated via a custom potential routine, which computes (in the frame of the star):
\begin{equation}
    \begin{aligned}
        d = &\sqrt{r^2 + r_{\rm BH}^2 - 2\,r\,r_{\rm BH}\,\sin\theta\,\sin\phi}\,,\\
        \Phi(r,\theta,\phi)
        = &-\frac{G\,M_{\rm BH}}{d}
           -\tfrac12\,\Omega_0^2\,r^2\sin^2\theta\\
           &+\frac{G\,M_{\rm BH}}{r_{\rm BH}^2}\,r\,\sin\theta\,\sin\phi\,,
    \end{aligned}
\end{equation}
where $(r,\theta,\phi)$ are positions in our simulation domain relative to the star, $r_{\rm BH} = a_\star$ is the separation between the star and the BH, and $\Omega_0$ is the orbital (co-rotation) frequency. The last term in this equation removes the constant and linear pieces of the companion's point-mass potential so that only the true tidal (quadratic and higher) variation remains.  During each hydro step, $\Phi$ is evaluated at cell centers and faces in all three spherical directions, its finite-difference gradient is applied as a source term to the momentum and energy equations, and the velocity is then updated semi-implicitly to include the exact Coriolis acceleration $-2\,\boldsymbol\Omega_0\times\mathbf v$, ensuring a self-consistent treatment of tidal and non-inertial forces (see \citealt{Stone2010} for a more detailed description).

\subsection{Accretion Disk}

In the frame of the black hole, the accretion disk is rotating with a velocity of 
\begin{equation}
    \vec v_{\rm disk} = \sqrt{\frac{GM_{\rm BH}}{r}} \hat{\theta},
\end{equation}
where $r$ is the distance to the black hole. If the black hole is stationary in the frame where a star is corotating with it at a distance of $r_\star$, the star has an angular velocity 
\begin{equation}
    \vec\Omega_{\star} = \sqrt{\frac{GM_{\rm BH}}{r_{\star}^3}} \hat{z}.
\end{equation}
The relative motion between the star and the accretion disk is produced via a moving disk in our simulation, with the star fixed at the center of the simulation domain. Hence, the accretion disk has a velocity of
\begin{equation}
\begin{aligned}
        \vec v_{\rm disk,rot} =\, &\vec v_{\rm disk} - \vec\Omega_{\star} \times \vec{r} = \vec v_{\rm disk} - \Omega_{\star}(-y \hat x + x \hat y)\\
        =\, &\Omega_{\star} y \hat x + \left( -\frac{z}{\sqrt{y^2+z^2}}\sqrt{\frac{GM_{\rm BH}}{r}} - \Omega_{\star} x\right)\hat y \\
        &+\frac{y}{\sqrt{y^2+z^2}}\sqrt{\frac{GM_{\rm BH}}{r}}\hat z
\end{aligned}
\end{equation}
Now transitioning to a frame centered on the star ($x',y',z'$), with the black hole located at ($x',y',z'$) = (0,$a_\star$,0) or ($r,\theta,\phi$) = ($a_\star$,$\pi/2$,$\pi/2$), 
\begin{equation}
\begin{aligned}
    x &= x'\\
    y &= y' - a_\star\\
    z &= z'\\
    r &= \sqrt{x'^2 + (y' - a_\star)^2 + z'^2}
\end{aligned}
\end{equation}

As in \citet{Yao2025}, we adopt a fiducial accretion disk with a Gaussian density profile and scale height $H_{\rm disk}$; we define $H_{\rm disk}$ as half of the Gaussian's full width at half maximum, consistent with the usual accretion disk definition of $H_{\rm disk}$ as the half-thickness of the disk. 
In the corotating frame, the accretion disk is initially tilted from the star--black-hole plane with $\theta_{\rm disk}\equiv {\rm arctan}(r_{\rm disk}/a_\star)$, where $r_{\rm disk}$ is the vertical height of the disk above the star along the $x'$ direction as indicated in Figure \ref{fig:schematic}, and $a_\star$ is the semi-major axis of the circular orbit.

\subsection{Stellar Orbit}

\begin{table}[htbp]
  \centering
  \caption{Disk and orbital parameters for the 3D Athena++ simulations of star-debris-disk interactions, for a $1M_{\odot}$ star orbiting a $10^6M_\odot$ SMBH on a circular orbit.}
  \begin{tabular}{@{}ccccc@{}}
    \toprule\toprule
    \multicolumn{5}{c}{\textbf{Model parameters}} \\
    \midrule
    $a/r_{\rm t}$ &
    $P_{\rm orb}\,(\mathrm{hrs})$ &
    $H_{\rm disk}\,(R_{\odot})$ &
    $v_{\star}\,(c)$ &
    $\rho_{\rm disk}\,(\mathrm{g\,cm^{-3}})$ \\
    \midrule
    $3.5$ & $18$ & \multirow{3}{*}{$5.6$} & $0.078$ & \multirow{3}{*}{$10^{-6}$} \\
    $5$   & $31$ &                        & $0.065$ &                          \\
    $8$   & $63$ &                        & $0.051$ &                          \\
    \bottomrule
  \end{tabular}
  \label{tab:params}
\end{table}

In our simulation, we place the star on a circular orbit around a $10^6M_{\odot}$ SMBH at three orbital separations (Table \ref{tab:params}):  $a/r_{\rm t} = 3.5,\, 5\,$ and $\,8$, where $\rt\equiv R_\star(\MBH/M_\star)^{1/3}$ is the tidal radius of the star; this corresponds to $a =$ 350, 500 and 800 $R_{\odot}$ 
and orbital periods of $P_{\rm orb} \sim 18,\,31,\,63\rm\, hrs$, respectively, for a $10^6 M_\odot$ SMBH and a solar type star. For convenience, we will refer to these three cases as ART3.5, ART5 and ART8, and the plots associated with them are separately colored green, blue and orange, respectively.

\section{Results} \label{sec:results}

\begin{figure*}
    \centering
    \includegraphics[width=\textwidth]{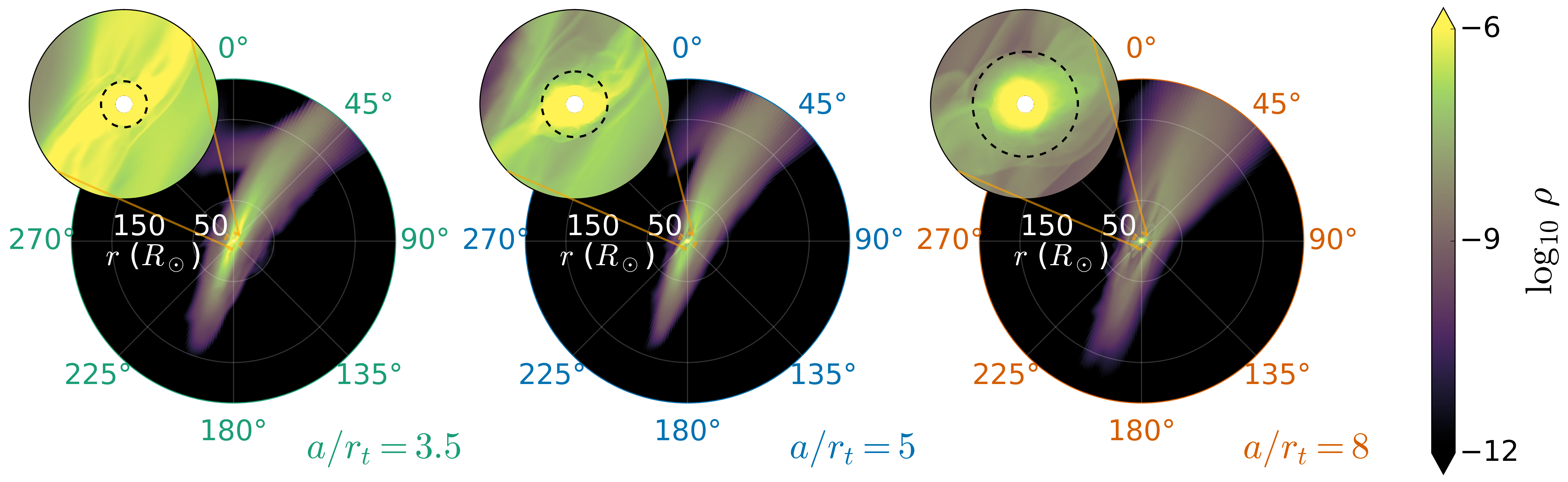} 
    \caption{Density contours of ART3.5 (\textit{left}), ART5 (\textit{center}), and ART8 (\textit{right}) for a $\theta$-slice in the orbital plane ($r-\phi$) showing the tidally stretched stellar debris stream roughly half an orbit after the previous disk collision. The inset plots show $r<10R_{\odot}$, where the size of the Hill sphere is shown via the dashed line. The debris stream evolves to cover a much larger volume for  larger $a/r_{\rm t}$, but can be significantly less dense, especially around the Hill sphere. The colorbar shared by these contours is capped at the disk's midplane density, highlighting areas of the stellar debris that are denser than the disk.}
    \label{fig:stream_contour}
\end{figure*}

\subsection{Mass Loss from Star}

When the star, surrounded by an extended stellar debris stream, collides head-on with the disk, most of the debris stream is immediately ablated. Subsequently, the collision compresses and heats up the surface envelope of the star, driving further mass loss that contributes to the formation of a new stellar debris stream over the course of half an orbit. 

In \citet{Yao2025}'s suite of star-disk collision simulations with physical conditions similar to those considered here, the mass stripped per encounter rises significantly after the first impact and saturates at a value $\sim\!20\times$ the first-collision yield around the fifth collision, corresponding to $\sim 10^{-5}-10^{-4}\,M_\odot$ per passage. The mass loss depends strongly on the disk scale height for the single-collision regime, but much more weakly for the multiple-collision regime because the disk primarily interacts with a dilute, shock-inflated stellar atmosphere that does not have time to cool efficiently before the next impact. 
For the multiple-collision case \citet{Yao2025} found that the mass stripped per collision is a function of the ratio of the ram pressure ($p_{\rm ram} = \tfrac12 \rho_{\rm disk} v_\star^2$) and the mean stellar pressure ($p_\star = G M_\star^2/(4 \pi R_\star^4) \simeq 10^{15} \, {\rm erg \, cm^{-3}}$), and the scale height of the disk,
with \begin{equation}
\begin{aligned}
\frac{\Delta M_{\star}}{M_{\star}} 
&\simeq 0.03 \left(\frac{p_{\rm ram}}{ p_\star} \right) \pfrac{H_{\rm disk}}{R_\star} \left[1 + \frac{H_{\rm disk}}{R_{\star}}\right]^{-1}\\
\label{eq:mstrip}
\end{aligned}
\end{equation}

\begin{figure*}
    \centering
    \includegraphics[width=\textwidth]{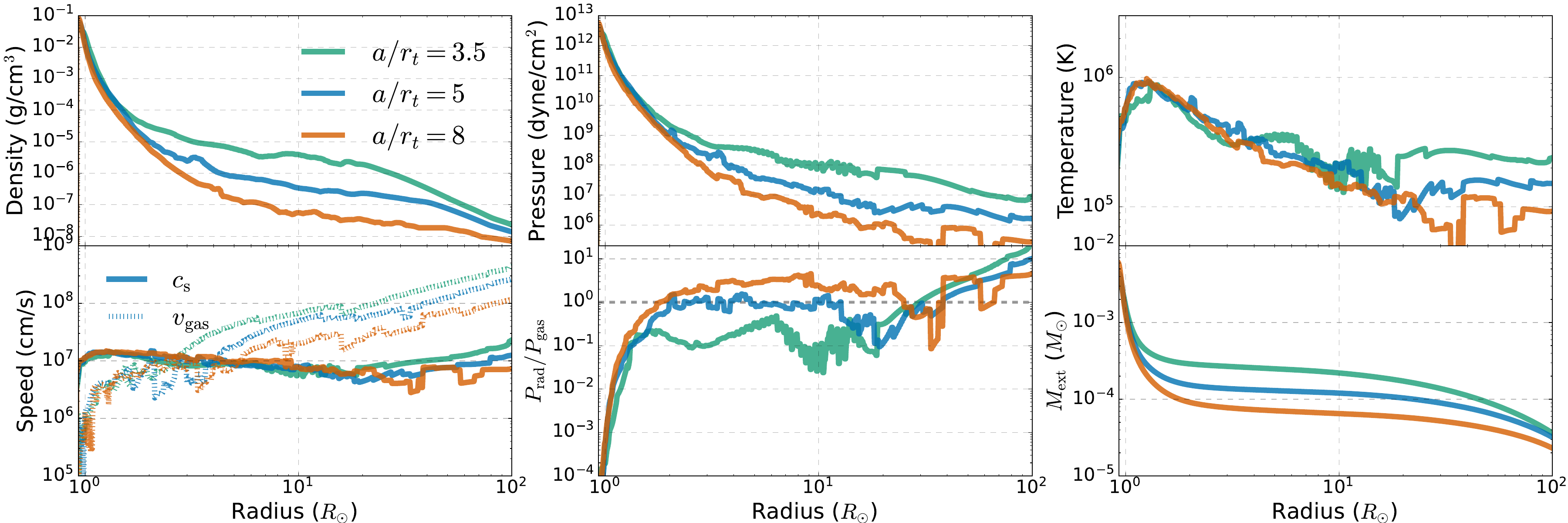}\\
    \caption{Stream density, pressure, temperature, sound speed, radiation-to-gas-pressure ratio profiles, and integrated exterior mass as a function of $r$ along the longest debris stream dimension ($R_1$) for ART3.5 (green), ART5 (blue), and ART8 (orange) after the puffy star experiences 3 disk collisions under tidal gravity (the same snapshot, immediately before the next collision, as the one shown in Figure \ref{fig:stream_contour}). The speed panel compares the total gas speed $v_{\rm gas}$ (dotted) to the local sound speed $c_{\rm s}$ (solid). At smaller $a/r_{\rm t}$, more material is stripped by the higher velocity collisions, and is denser along the stream and inside the Hill sphere.}
    \label{fig:profile} 
\end{figure*}

Direct comparison between this fitting function and our simulation results here is complicated by the inclusion of the accretion disk's Keplerian rotation, which increases both the magnitude of the collision velocity and the effective scale height experienced by the star by $\sqrt 2$. Nevertheless, we find mass loss consistent with Equation \ref{eq:mstrip} when taking these factors into account. Using Equation \ref{eq:mstrip} with model parameters listed in Table \ref{tab:params} results in $1.4\times10^{-4} M_\odot$, $9.7\times10^{-5} M_\odot$ and $6.0\times10^{-5} M_\odot$ stripped per collision for ART3.5, ART5 and ART8, respectively. In the simulations presented in this paper, we find that the mass loss from the 3rd collision (after introducing the BH's potential) is $2\times10^{-4} M_\odot$, $1\times10^{-4} M_\odot$ and $9\times10^{-5} M_\odot$ for ART3.5, ART5 and ART8, respectively. Consequently, the inclusion of the tidal gravity, the tidally stretched debris stream, and the change in the collision geometry do not have a notable influence on the stellar mass loss per collision for the models tested here.

\subsection{Stellar Debris Stream}

\begin{figure}
    \centering
    \includegraphics[width=\columnwidth]{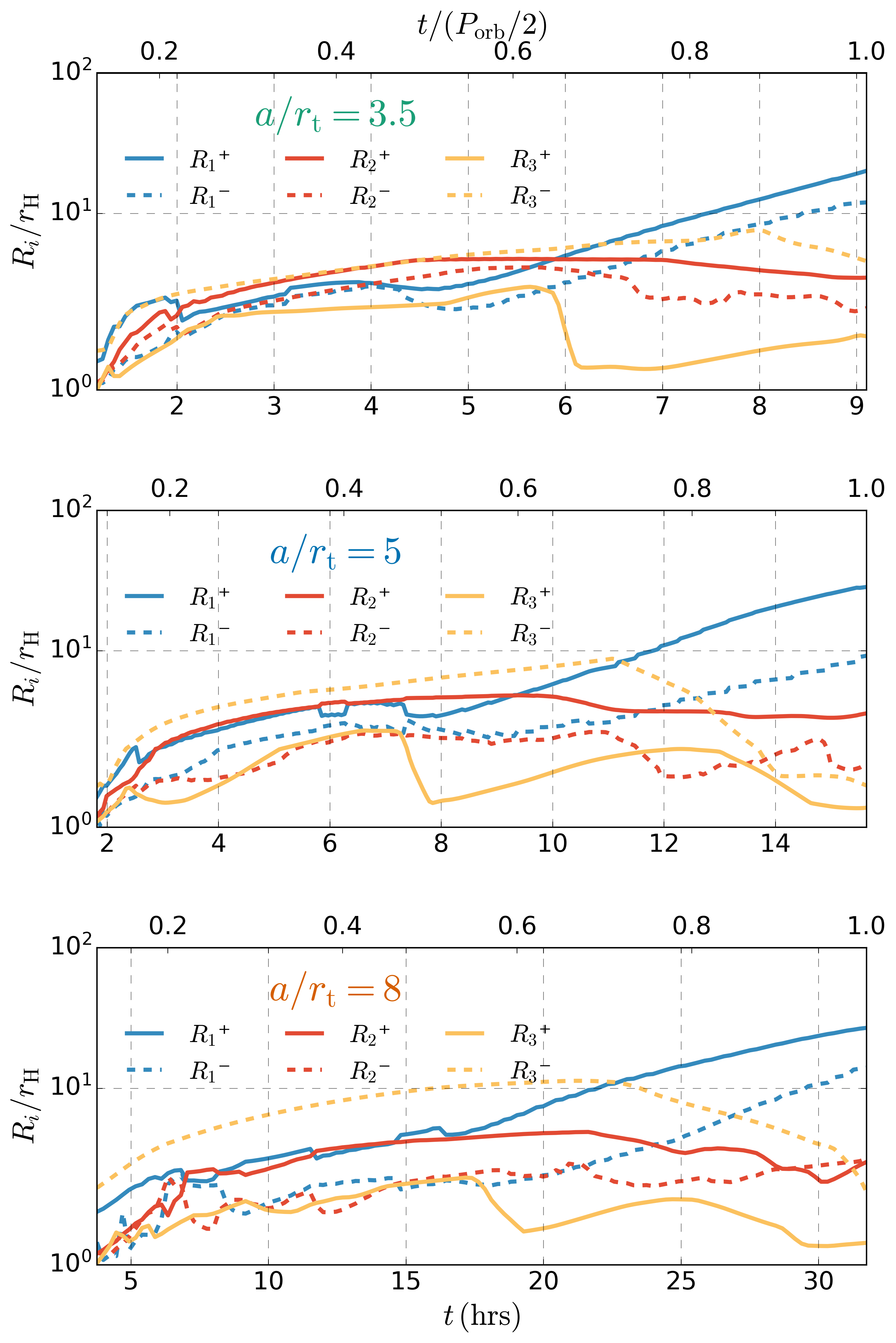}
    \caption{Evolution of the 3 pairs of stellar debris stream axes that trace the 90\% mass length along each direction, normalized by the size of the Hill sphere $r_H$. $+$ and $-$ axes are parallel to each other and extend from the star in two opposite directions. $R_1$ and $R_2$ are in the orbital plane, where the former is the longest axis and sets the collision duration. The debris stream evolution is nearly independent of $a/r_t$ when normalized to the size of the Hill sphere.}
    \label{fig:stream} 
\end{figure} 

\citet{Yao2025} proposed that QPE flares are powered by collisions between stellar debris streams and the accretion disk, as they are more likely to give rise to more shocked mass, longer flare duration, and more total shocked debris energy than direct star-disk collisions alone. The stream originates from previous star-disk collisions where the debris quickly exits the Hill sphere and then evolves ballistically in the BH potential, where tidal shear stretches it into a long, thin stream. The spread in semimajor axis $\Delta a/a \sim \mathcal{O}(v_{\rm esc}/v_\star)$ sets the stream's in-plane major axis ($R_1$), while the dispersion in the $x$-component of angular momentum fixes the in-plane minor axis ($R_2$). \citet{Linial2025} present a more detailed calculation of the debris dynamics outside the Hill sphere using Hill's equations. This results in the evolution of the 3 stream dimensions as a triaxial ellipsoid: $R_1$ and $R_2$ lie in the orbital plane, and $R_3$ is perpendicular to that plane. This geometry is critical for the resulting debris-disk interaction since it sets both the flare duration (via the disk crossing time $t_{\rm dur}\sim 2R_1/v_{\star}$) and the shocked disk mass via the collision cross section ($R_2$ and $R_3$). In this paper, we improve upon these previous estimates by including the tidal potential of the black hole in our 3D simulations. 

In Figure \ref{fig:stream_contour}, we show the density of the stellar debris stream roughly half an orbit after the previous impact and immediately before encountering the disk again for ART3.5, ART5, and ART8. The stream is asymmetric, most significantly along $R_1$ (the longer axis in the orbital plane shown), which is caused by the asymmetric ejection of stripped material from the side of the star experiencing a head-on disk collision (stronger) versus the opposite side (weaker). For smaller $a/r_{\rm t}$, the stream is denser and less extended. The Hill sphere (dashed black circle) is almost filled with dense gas. In contrast, larger $a/r_{\rm t}$ runs show a more dilute and extended stream, where the density of material in the outer parts of the Hill sphere can be nearly indistinguishable from stream material in its vicinity. 

A more quantitative depiction in Figure \ref{fig:profile} shows the density, pressure, temperature, sound speed, radiation-to-gas-pressure ratio, and exterior mass profiles along $R_1$ in the streams. In every case, material in the stream, once exiting the Hill sphere, is moving supersonically (comparing the total gas speed $v_{\rm gas}$ to the sound speed $c_{\rm s}$), is radiation dominated, and has a temperature around $\sim10^5$ K. The stream's peak density is comparable to the disk mid-plane density ($10^{-6}$ g cm$^{-3}$) out to progressively larger radii as $a/r_{\rm t}$ decreases: within $\sim 30\,R_{\odot}$ for ART3.5, $\sim 5\,R_{\odot}$ for ART5, and only $\sim 3\,R_{\odot}$ for ART8. Notably, for ART3.5 the region denser than the disk is not a roughly spherical $\sim 30\,R_{\odot}$ volume but a more cylindrical region elongated along the stream's long axis, which can prolong the interaction timescale with the disk. This partially sets the amount of disk mass that is strongly shocked by the stellar debris, which therefore increases with decreasing $a/r_{\rm t}$, being about a factor of $\sim 3$ larger in ART5 than in ART8, and a further factor of $\sim 3$ larger in ART3.5 than in ART5. Overall, the profiles in Figure~\ref{fig:profile} obey the self-similarity found in \citet{Linial2025} reasonably well, with the primary difference between the runs being the lower overall debris mass and mean ejecta density at larger $a/r_{\rm t}$, consistent with the self-similar scaling of \citet{Linial2025}.

In Figure \ref{fig:stream}, we track the evolution of $R_1$, $R_2$, and $R_3$ by taking the 90\% mass length of the highest mean density axis in-plane, the in-plane axis perpendicular to it, and the perpendicular polar axis, respectively. We further distinguish each axis into two separate parallel halves $R_i^+$ and $R_i^-$, to capture asymmetries in stream dimensions. The main cause of asymmetries, particularly along $R_1$ and $R_3$ originates from the collision velocity vector along $\vec v_{\star}$ and $\vec v_{\rm disk}$, respectively, that elongates the stream extending from the head-on collision side due to higher ejecta velocity. Like in \citet{Linial2025}, we see the orbital focusing along $R_3$ predicted by the ballistic Hill's solution that starts around $t/(P_{\rm orb}/2) = 0.7$ and is most compressed around $t/(P_{\rm orb}/2) = 1$, but as expected the pressure gradient in our simulation halts the collapse. Finally, when normalized by $r_{\rm H}$, the stream evolution for the three simulations is qualitatively and quantitatively nearly indistinguishable. 

\subsection{Energetics of Shocked Disk and Debris Stream}

\begin{figure*}
    \centering
    \includegraphics[width=\textwidth]{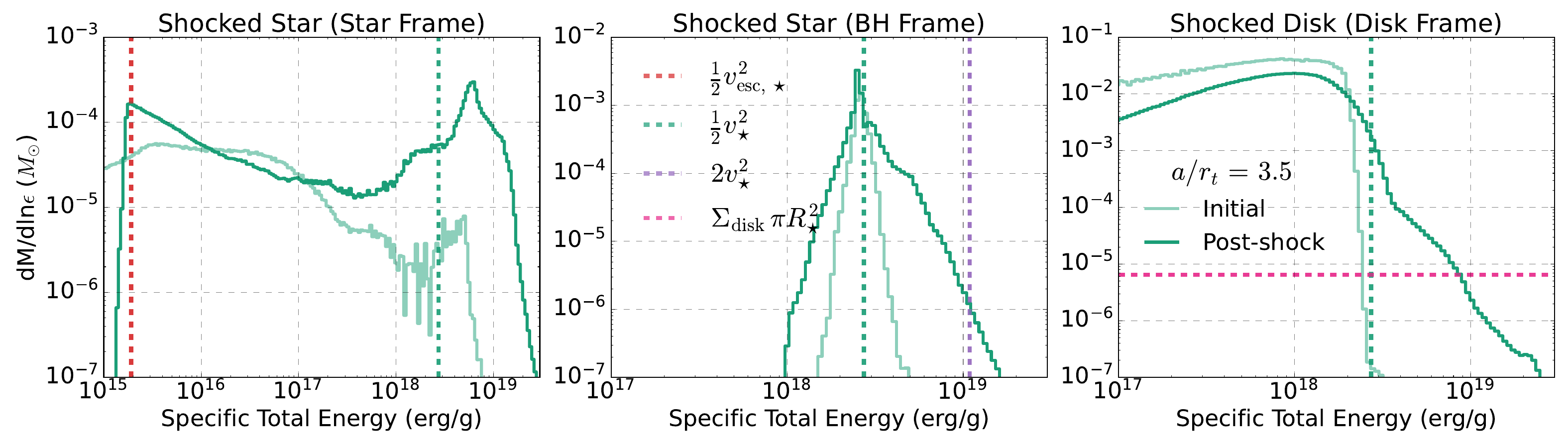}\\
    \includegraphics[width=\textwidth]{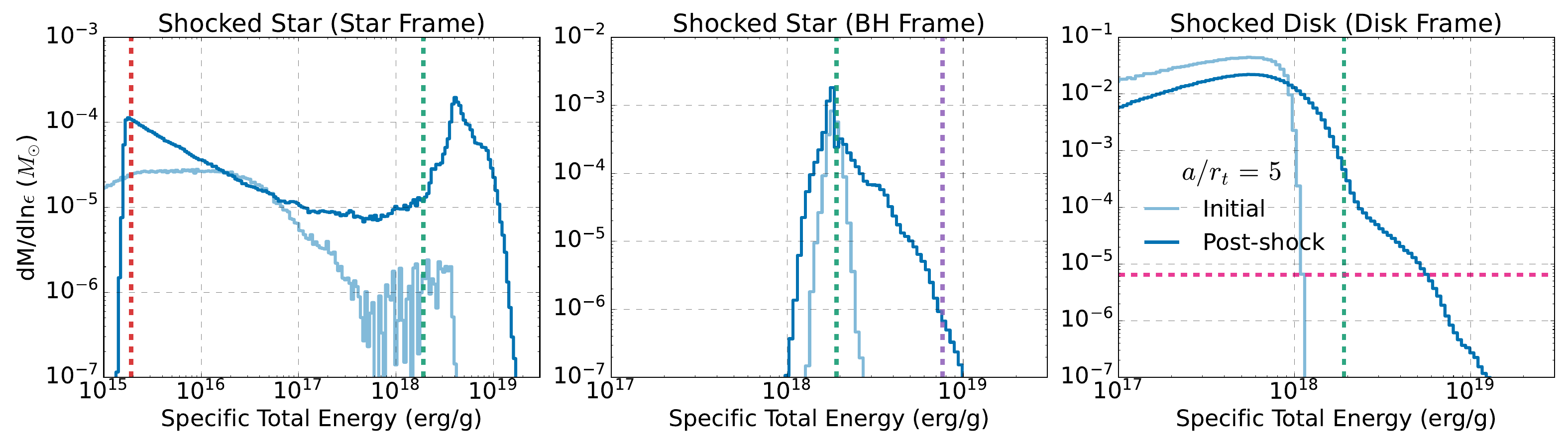}\\
    \includegraphics[width=\textwidth]{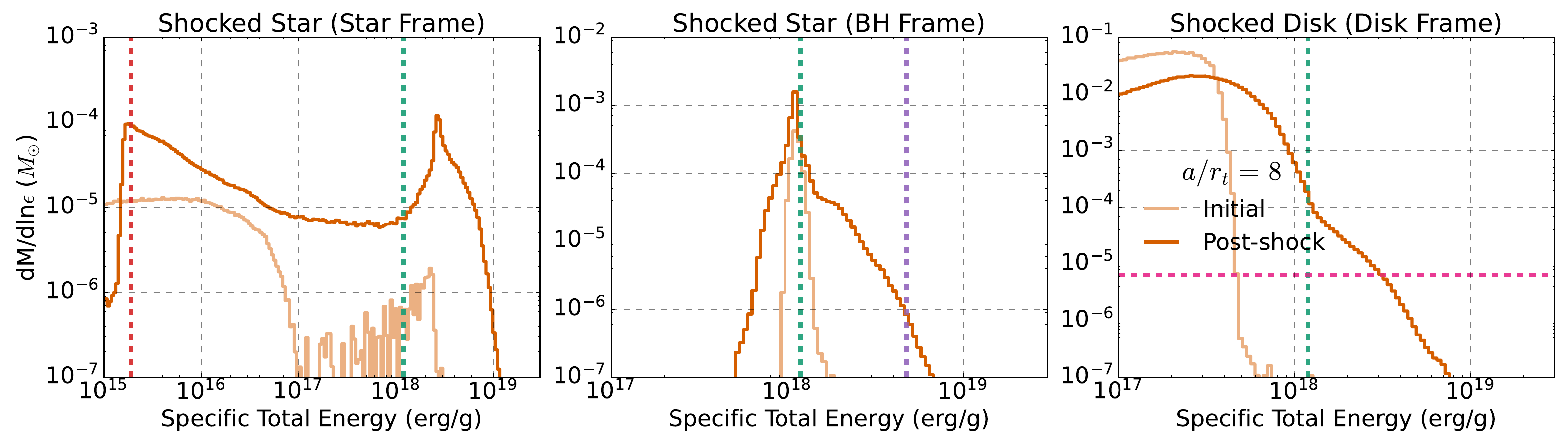}\\
    \caption{Energy distribution of stellar debris in the frame of the star (\textit{left}) and the frame of the black hole (\textit{center}) and disk material in the frame of the disk (\textit{right}) before (lighter shade) and after (darker shade) a collision for ART3.5 (\textit{top}), ART5 (\textit{middle}), and ART8 (\textit{bottom}). The stellar debris has a typical energy set by the stellar escape velocity $v_{\rm esc,\star}$, whereas shocked star and disk material have specific energy set by the stellar orbital velocity $v_\star$. We categorize material as shocked if its specific energy $\epsilon_{\rm tot}>\frac{1}{2} v_{\star}^2$, and unbound from the black hole if $\epsilon_{\rm tot, BH}>2 v_{\star}^2$. In the \textit{right} panels, the energy distribution of undisturbed disk material broadens due to numerical diffusion, which increases the uncertainty of the measured shocked disk mass.}
    \label{fig:energy}
\end{figure*}

\begin{figure*}
    \centering
    \includegraphics[width=\textwidth]{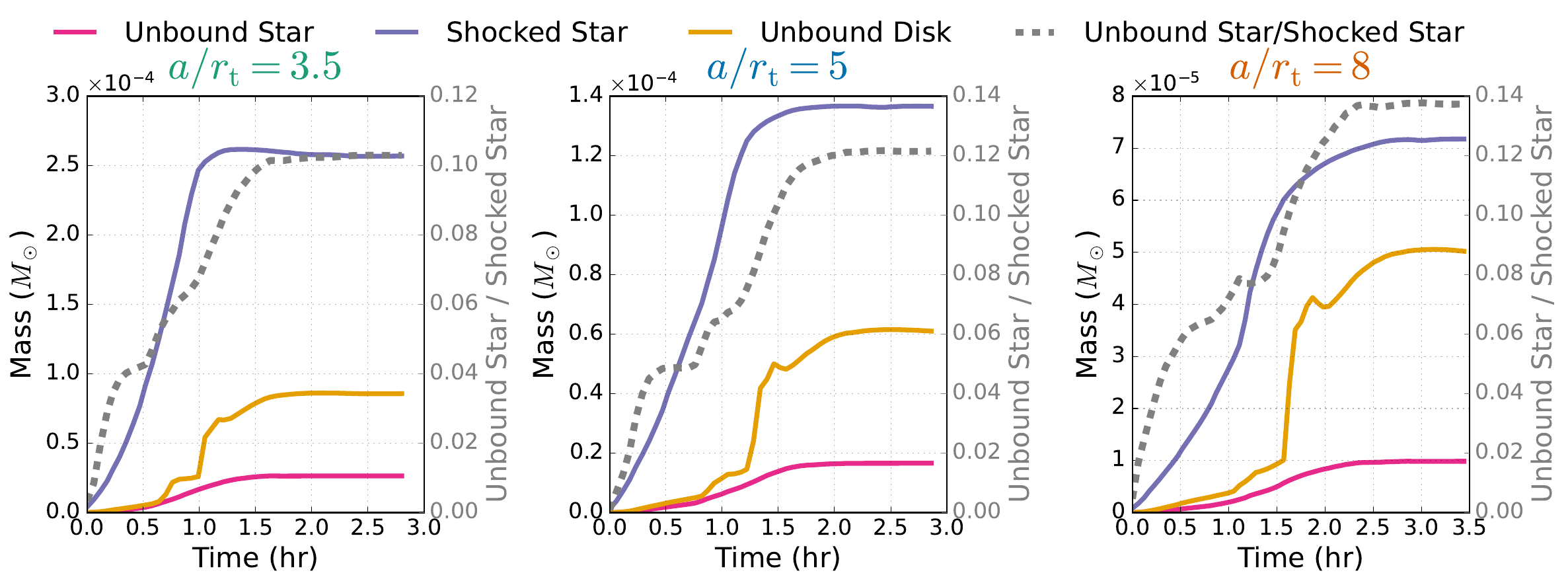}\\
    \caption{Amount of star and disk material unbound from the black hole for ART3.5 (\textit{left}), ART5 (\textit{center}), and ART8 (\textit{right}). The unbound mass from the star is also plotted as a ratio (gray dashed lines for the right x-axes) to the amount of shocked stellar material ($\epsilon_{\rm tot}>\frac{1}{2} v_{\star}^2$). Shocked disk mass is not plotted due to uncertainties arising from numerical diffusion.}
    \label{fig:unbound}
\end{figure*}

\begin{table}[htbp]
  \centering
  \caption{Star and disk material stripped or unbound from the black hole following a collision (distinguished in the simulations by their corresponding passive scalars, and denoted here by ``$\star$" and ``disk" subscripts). These quantities are measured by comparing their total specific energy in the star's frame ($\epsilon_\star$) or in the black hole's frame ($\epsilon_{\rm BH}$) with other key parameters such as the specific gravitational binding energy or specific kinetic energy derived from the star's Keplerian orbital velocity. Masses are listed in units of $M_{\odot}$.}
  \begin{tabular}{@{}ccccc@{}}
    \toprule\toprule
    $a/r_{\rm t}$ &
    $v_{\star}\,(c)$ & 
    $M_{\rm \star, shocked}$ &
    $M_{\rm \star, unbound}$ &
    $M_{\rm disk, unbound}$ \\
    \midrule
    $3.5$ & $0.078$ & $3\times10^{-4}$ & $3\times10^{-5}$ & $9\times10^{-5}$ \\
    $5$   & $0.065$ & $1\times10^{-4}$ & $2\times10^{-5}$ & $6\times10^{-5}$ \\
    $8$   & $0.051$ & $7\times10^{-5}$ & $1\times10^{-5}$ & $5\times10^{-5}$ \\
    \bottomrule
  \end{tabular}
  \label{tab:masses}
\end{table}
In Figure \ref{fig:energy}, we plot the energy distribution for material in our simulation domain, traced with the star and disk passive scalars, for a range of specific total energy, defined as the sum of local specific kinetic and thermal energies. The lighter color represents the unshocked stellar stream and disk in a simulation snapshot before another imminent collision, whereas the darker color represents the energy of shocked material as their corresponding shocked mass peaks (this differs in time for star vs disk). 

\begin{figure*}
    \centering
    \includegraphics[width=\textwidth]{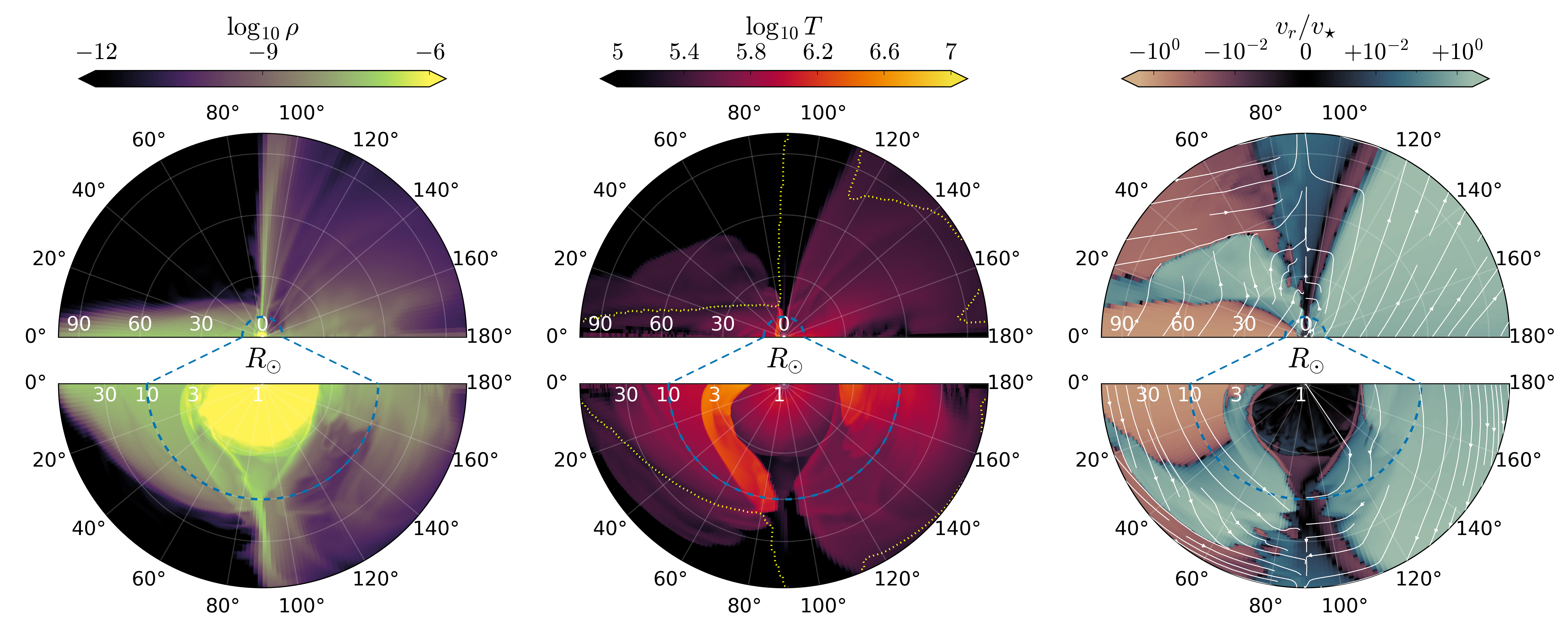}\\
    \caption{From \textit{left} to \textit{right}, density, temperature, and radial velocity contours of a simulation snapshot for ART5. The two hemispheres are the same data mirrored and plotted linearly (\textit{top}) and logarithmically (\textit{bottom}) in radius $r$ to resolve both large- and small-scale structures, respectively. The dashed line indicates the same spatial region within $r<10 R_{\odot}$ in each plot. The \textit{right} column shows the radial velocity of material in the star's frame. The snapshot is taken during the collision when the star sits approximately at the center of the disk, and a wind-like outflow is produced by the shocked debris-disk mixture traced by the white streamlines in the \textit{right} column. The yellow dotted line in the temperature panel indicates the position where $\tau\sim c/v_\star$, which is viewing-angle dependent. Some angles can see material shocked to temperatures as high as $\sim 10^{6}K$, capable of producing the signature soft X-ray flares seen in QPEs. Results are similar for ART3.5 and ART8.}
    \label{fig:6contours}
\end{figure*}

Stellar-debris energies are calculated in the star's frame (left panel) and in the black hole's frame (middle panel), counting only stellar material no longer bound to the star. This snapshot is chosen roughly when half of the stellar stream has been shocked by the disk ($R_1^+$), which is also the time shown in Figure \ref{fig:6contours}.  In the star frame, we see that most of the stellar debris liberated in the previous collision is shocked to a specific energy $\gtrsim \frac{1}{2}v_\star^2$. In the black hole's frame, as expected, most stellar debris has a specific energy of $\sim \frac{1}{2}v_\star^2$, yet a few percent of the shocked debris exceeds $2v_\star^2$ and is therefore unbound from the black hole. 

Disk energies are evaluated in the disk frame (right panel) by subtracting the mean specific total energy of an undisturbed disk from all disk material after the collision. This snapshot was chosen after the stellar Hill sphere has crossed the entire disk, creating a large bubble of shocked disk material from the shock breakout (see Fig. \ref{fig:schematic}). Numerical diffusion naturally increases the average specific total energy for the disk moving in our simulation domain. Opting for a thicker disk has reduced the effect, but still introduces an $\lesssim 50\%$ error in the measured shocked disk mass at this snapshot (measured in a separate convergence test not shown in Fig. \ref{fig:energy}). Nevertheless, shocked disk material has a characteristic specific energy $> \frac{1}{2}v_\star^2$, and this is evident in the energy distribution as the slope change occurs at this energy threshold. The shocked disk mass is $\sim 3-4$ times lower in ART8 than in ART5, with the latter a further $\sim 3-4$ times lower than ART3.5, but all higher than the naive estimate of $\sim \Sigma_{\rm disk}\pi R_\star^2$ based on the unshocked size of the star. This difference originates from the fact that the effective cross-section for stellar material within the Hill sphere at a density above that of the disk is $\sim (3R_{\odot})^2$ for ART8, $\sim (5R_{\odot})^2$ for ART5, and even larger for ART3.5 (albeit less spherically symmetric). We therefore expect this trend to continue with increasingly larger $a/r_{\rm t}$, until the effective cross section drops to $\sim R_{\odot}^2$. 

To address how much material is unbound from the system, both originating from the star and the disk, we perform a more detailed analysis in Figure \ref{fig:unbound} by summing cell by cell the stellar and disk material with energy exceeding the gravitational energy of the black hole at that cell location, over the duration of the third collision. For star material, about $10-14\%$ of the shocked mass is unbound; for disk material, while the unbound mass is similar for ART5 and ART8, the collision in ART3.5 unbinds a meaningfully higher amount of mass due to more extended dense material along the debris stream that exceeds even the size of the Hill sphere; this trend can likely be extrapolated to even larger $a/r_{\rm t}$, where the unbound disk mass is set by the disk column swept up by the cross section of the puffy bound star, so long as the dense stellar material remains contained within the Hill sphere. Observationally, unbound star and disk material could manifest themselves as the outflow observed in GSN 069 by high-resolution X-ray spectroscopy carried out by \citet{Kosec2025}, which is more highly ionized during the QPE-emitting phase than during quiescence. However, the observed velocity from the Doppler effect is $\sim 1700-2900 \, \rm km/s$, significantly slower than that of the unbound mass seen in our simulations. While the observed outflow velocities are lower than the fastest unbound material in our simulations, the unbound disk and stellar debris ejected per collision may nonetheless contribute to the fast outflows inferred in some TDEs, as discussed by \citet{Linial2026}.

Figure \ref{fig:unbound} shows that a significant amount of disk mass can be unbound during the debris-disk collisions ($\sim5\times10^{-5}-10^{-4}M_{\odot}$ per collision).   This could have implications for the lifetime of the disk if it has a finite mass, such as a newly formed TDE disk of $\sim 0.1-0.5 M_\odot$.  On the other hand, most of the stripped stellar material remains bound to the system after each collision, and can replenish the disk (and the stripped stellar mass per collision exceeds the shocked disk mass). \citet{Linial2024b} discuss this coupled star-disk evolution, in which stripped stellar material is gradually
incorporated into the disk. Our star-centered simulations lack the ability to model this replenishment, which could be interesting to explore in future global simulations. 

\subsection{Flare Duration from Shocked Debris Stream} \label{sec:duration}

\begin{figure}
    \centering
    \includegraphics[width=\columnwidth]{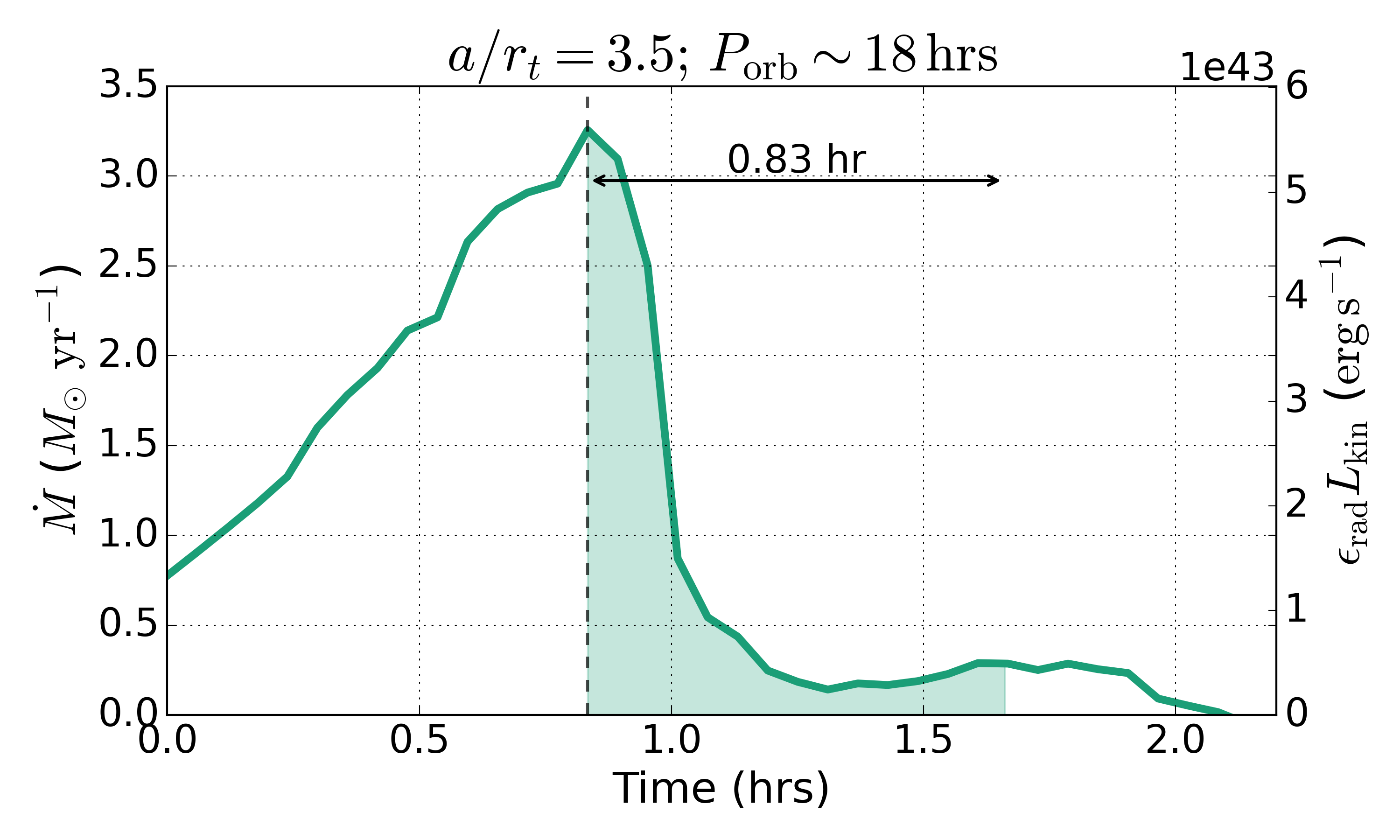}\\
    \includegraphics[width=\columnwidth]{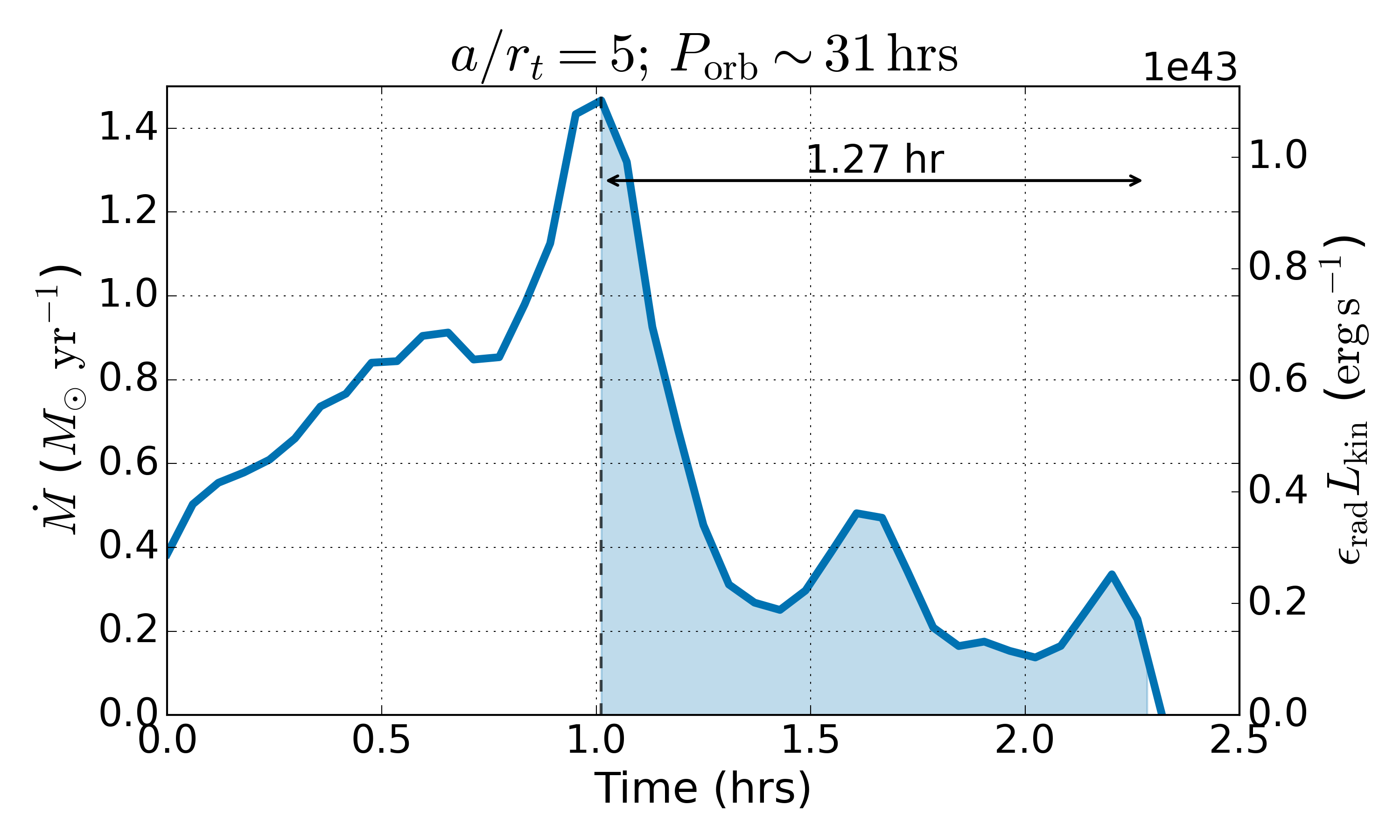}\\
    \includegraphics[width=\columnwidth]{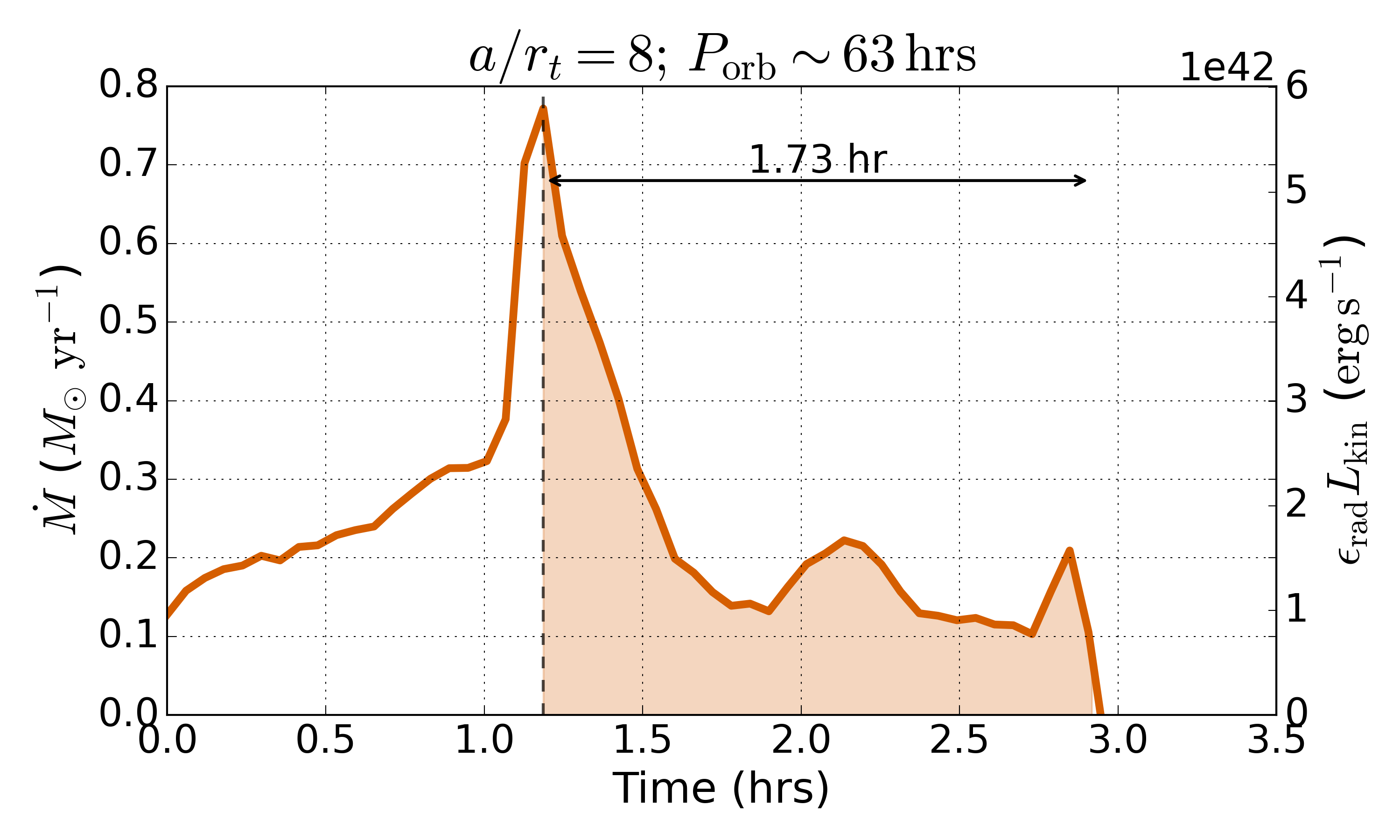}\\
    \caption{Shocked stellar debris mass over time for ART3.5 (\textit{top}), ART5 (\textit{middle}), and ART8 (\textit{bottom}). On the right axes, an estimate of the bolometric luminosity is converted from $\dot M$ assuming radiative efficiency  $\epsilon_{\rm rad} \sim 0.1$. Only the post-peak duration is taken into account since the pre-collision stream is truncated artificially in our simulations; we estimate the total flare duration to be three times the post-peak duration (see discussion in \S \ref{sec:duration}). The measured flare duty cycle ($\sim10\%$, assuming 1 flare per orbit) and luminosity broadly reproduce the observed properties of QPEs. }
    \label{fig:masslc}
\end{figure}

Our simulations allow us to directly track the amount of shocked mass over the entire course of the stream-disk interaction. To trace the shocked stellar material ($\epsilon_{\rm tot} > \frac{1}{2}v^2_{\star}$ in the frame of the star), we measure the total amount of stellar mass shocked within and leaving the simulation domain, and calculate the corresponding $\dot M$ as plotted in Figure \ref{fig:masslc}. The exact luminosity produced by the shocked mass is uncertain and depends on the radiative processes, which will be explored in future work. To guide the reader, the right axes of Figure \ref{fig:masslc} show $L =\frac{1}{2}\epsilon_{\rm rad}\dot M v_{\star}^2$ with radiative efficiency $\epsilon_{\rm rad}\sim 0.1$. We return to this in the next section. 

To quantify the duration of the ``flare", we rely on the length of time (shaded region) after $\dot M$ peaks (vertical dashed line) and drops to $\sim 10\%$ of the peak value, which essentially tracks the length of $R_1^-$. This is because we artificially truncate $R_1^+$ when initializing the disk (to optimize for resolution and minimize numerical diffusion). Since $R_1^+$ is roughly $2-3\times$ the length of $R_1^-$ without truncation (see Fig. \ref{fig:stream}), the total ``flare" duration should be at least $\sim 3R_1^-/v_\star$.

Our estimated flare durations are then $\sim 3\times0.83\,\rm hr = 2.49\, hrs$, $\sim 3\times1.27\,\rm hr = 3.81\, hrs$ and $\sim 3\times1.73\,\rm hr = 5.19\, hrs$, corresponding to flare duty cycle of $\sim 16-27\%$ for the two-flare-per-orbit case, and $\sim 8-14\%$ for the one-flare-per-orbit case. As explored in \citet{Linial2025}, this duty cycle corresponds to the debris stream crossing time through the accretion disk, which scales with $a/r_{\rm t}$ and hence $r_{\rm H}$, as well as the Keplerian orbital velocity via
\begin{equation}\label{eq:duration}
    t_{\rm dur,s} \sim 2R_1/v_{\star}\simeq  2.7 \, {\rm hours} \, P_{\rm orb,day}M_{\star,\odot}^{1/3} M_{\rm BH,6}^{-1/3}, 
\end{equation}
where we have normalized the constant in the front of Equation \ref{eq:duration} using our simulations.  Variations in the observed QPE duty cycle at a given orbital period can be partially attributed to variations in stellar mass and host black hole mass. 

In principle, we can directly measure the amount of disk material shocked by the star and the debris in the Hill sphere. For the disk thickness we consider, this duration would be significantly shorter than Equation \ref{eq:duration}, as it scales directly with the disk crossing time of the star when the disk is thicker than the part of the stream denser than the disk (true for ART5 and ART8):
\begin{equation}\label{eq:duration_disk}
    t_{\rm dur,d} \sim 2H_{\rm disk}/v_{\star} \simeq 0.018 {\rm hr} \,H_{\rm disk,R_\odot} P_{\rm orb, day}^{1/3}M_{\rm BH,6}^{-1/3}.
\end{equation}
Here, the duration is much shorter than the flare produced by the stellar debris and this disk-crossing timescale also would not produce the observed linear correlation between duration and orbital period observed across QPE systems. We further note that this hydrodynamic crossing time is much shorter than the
photon diffusion time out of the shocked disk column, so any flare powered by the shocked disk would be broadened to the diffusion timescale rather than tracking $t_{\rm dur,d}$; the disk-crossing time therefore cannot accurately set the observed flare duration. On the other hand, for ART3.5, the dense stream extends to about $60 R_{\odot}$ in length along $R_1$, significantly lengthening the interaction timescale with the disk relative to equation \ref{eq:duration_disk}.   We choose not to present quantitative properties of the shocked disk energetics analogous to Figure \ref{fig:masslc} because of concerns that numerical diffusion modestly changes the disk properties as it moves across the grid; future higher resolution simulations or careful AMR modeling would remedy this and would be useful for fully quantifying the shocked disk properties.

\subsection{Emission from Shocked Debris Stream} \label{sec:emission}

In the model of star-debris-disk interactions, two sources can power QPE emission: debris shocked by the accretion disk and the accretion disk shocked by the star and the Hill sphere. The latter case is more difficult to follow in our current simulation centered on the frame of the star, and our results for the shocked disk do not differ significantly from those of \citet{Yao2025}. The dominant effects here are controlled primarily by $a/r_{\rm t}$, which sets the collision cross-section: assuming the same disk parameters, more disk mass is shocked for smaller $a/r_{\rm t}$. We refer the reader to the discussions in \citet{Linial2025} and \citet{Huang2025} for analytic and numerical estimates, respectively. In this section, we focus on the stream-disk collision case which differs significantly from previous  calculations that neglect the effects of the SMBH's tidal potential. 

In Figure \ref{fig:6contours}, we show the density, temperature, and radial velocity contours of a collision snapshot when the star is around the midplane of the disk for ART5 (snapshots for ART3.5 and ART8 are qualitatively similar). The \textit{top} and \textit{bottom} hemispheres are renderings of the same region above the disk in the direction where the stream is, but the included quantity is plotted linearly (\textit{top} hemisphere) and logarithmically (\textit{bottom} hemispheres) in the radial direction to isolate larger and smaller scale differences, respectively. It can be seen in the \textit{left} and \textit{right} panels that half of the stream that trails the star remains undisturbed at this time (along $R_1^-$), while the other half is shocked by the rotating disk to produce a wedge-shaped outflow filled by a mixture of shocked stellar and disk material. On the \textit{right} panel, material in the wind-like outflow has a nearly constant-velocity, roughly at the orbital velocity of the star; this is also the dominant component of the shocked stellar debris in mass, deflected at an angle due to disk rotation. 

The \textit{middle} panel of Figure \ref{fig:6contours} shows that the shocked stellar debris has a range of temperatures from $\sim 6\times10^5 K$ along the outflow, to $\sim 3\times10^6 K$ near the base of the stream (these temperatures are significantly higher than in \citealt{Yao2025}, because the density of the stellar debris is much lower when it is tidally stretched by the black hole). 
On the same panel, we overplot the diffusion radius (where $\tau\sim c/v_\star$) by radially integrating over the density of the material from our outer boundary inwards, as indicated by the yellow dotted lines. Material along the stream and the shocked outflow to the right of the stream are optically thick, resulting in lower-temperature emission. To the left of the stream, the accretion disk is less optically thick, pushing the diffusion radius closer to the base of the stream where the temperature is much higher. The temperatures inferred in this figure are broadly consistent with the blackbody fit values $kT\sim 50-200\rm eV$ reported for many QPE sources. 

As discussed in the previous section, even when assuming a radiative efficiency of $\epsilon_{\rm rad} = 0.1$, the resulting flare bolometric luminosity, as plotted in the secondary axis in Figure \ref{fig:masslc}, can reach $\sim 10^{42}-10^{43} \rm erg s^{-1}$, consistent with estimates from QPE observations. Smaller $a/r_{\rm t}$ runs have higher luminosity than larger $a/r_{\rm t}$ runs because of more shocked mass and higher collision velocity. It is worth noting that the bolometric luminosity is likely a sublinear function of the kinetic luminosity, set by the mass content of the stream and the optical depth of the collision front. As a result, while Figure \ref{fig:masslc} shows a sharp peak in the shape of the $\dot M$ curve, the observed flare profile can be flatter in reality (see detailed discussion in \citealt{Linial2025}'s sections 4.3.1, 4.3.2, and 5.2). To produce a more accurate light curve that reproduces the flare shape and spectral evolution, detailed radiative transfer physics needs to be included, which we defer to future work. 

\section{Summary and Discussion} \label{sec:discussion}
\subsection{Summary}

Hydrodynamic simulations of repeated star-disk collisions in \citet{Yao2025} showed that quasi-periodic eruptions (QPEs) are most likely to be produced by stellar debris liberated in previous collisions impacting the accretion disk half an orbit later (a direct star-disk collision alone involves much less shocked mass and energy and cannot explain longer-period QPE energetics; e.g., \citealt{Mummery2025,Guo2026a}).   In this paper, we have incorporated SMBH tidal gravity and disk Keplerian rotation in 3D \texttt{Athena++} calculations of star-disk collisions to self-consistently follow the dynamics of the stellar debris including the SMBH's gravity.   We find that the debris stripped from the star in a star-disk collision exits the Hill sphere and is sheared into an extended, asymmetric stream (Fig. \ref{fig:stream_contour}) whose geometry (axes $R_{1,2,3}$) is set primarily by orbital dynamics; when scaled to the Hill radius, the stream evolution (Fig. \ref{fig:stream}) is nearly identical for the three different orbits we have simulated ($a/r_{\rm t}=3.5,\,5\,,$ and $8$). 

The resulting encounter between the stellar debris stream and accretion disk shocks both the debris and some disk gas to specific energies $\gtrsim \frac{1}{2} v_\star^2$ (Fig. \ref{fig:energy}), and produces a wind-like outflow of debris due to the extended time duration of the collision. For most of the conditions we simulate (a solar mass star and orbital periods of roughly a day for a $10^6 M_\odot$ BH), the stellar debris density is less than the disk density, so the dominant high specific energy shocked material is the stellar debris shocked by the denser disk. This ratio of shocked disk to shocked stellar debris energy is, however, a strong function of orbital period: the stream density (and thus the effective collision cross section with the disk) increases at smaller orbital periods (smaller $a/r_{\rm t}$), yielding substantially more shocked disk mass in the shortest-orbit case (ART3.5), for which the two shocked components have similar mass. 

Our simulations do not include radiation transfer and so cannot model the resulting flare light curve or its spectrum. As a proxy for some flare properties, we measure the shocked stellar mass and energy as a function of time in the simulations (Fig. \ref{fig:masslc}); this allows us to estimate the flare duration and provide an upper limit on the luminosity and energy radiated by the shocked debris.   We infer flare energetics consistent with observations and flare durations set by the time for the stream to impact the disk $\sim R_1/v_\star$ (where $R_1$ is the long axis of the stellar debris and $v_\star$ is the star's orbital velocity).   This produces duty cycles of order $\sim 10-20\%$ for the flare, depending on model assumptions such as the number of observable flares per orbit.

This ratio of flare duration to recurrence time is of order that observed, as argued analytically in \citet{Linial2025}. Overall, SMBH tides prolong the stream-disk interactions and naturally connect QPE durations to the orbital period. To the best of our knowledge, this is one of the few  QPE models that are able to explain this remarkable regularity of the observed duty cycle in QPEs.

\subsection{Emission mechanism and the flare timing pattern}
The detailed morphology of QPE flares (e.g. sometimes fast-rise/slow-decay, sometimes symmetric) likely reflects radiative transfer through a rapidly evolving, radiation-dominated collision front not captured in our pure-hydro simulations. In the stream-disk picture, the collision naturally drives a wind-like outflow containing mixed disk and debris material; the observed soft X-ray emission then arises from photon-starved shocks and subsequent reprocessing in the expanding outflow \citep{Linial2023, Linial2025}. Our simulations show debris-disk collisions can shock material at the base of the stream to reach temperatures as high as $\sim 3\times 10^6\,{\rm K}$ (Fig. \ref{fig:6contours}), capable of producing soft-X-ray photons. Interpreting these temperatures as a blackbody, however, implicitly assumes efficient thermalization (rapid photon production and equilibration) at the emitting depth. Under this assumption, the inferred blackbody temperature corresponds to a lower limit on the characteristic photon energy; if the shock is photon-starved, incomplete thermalization can yield a harder spectrum even at comparable energy densities. Meanwhile, most of the shocked material at the thermalization depth lies beneath the diffusion radius, so the emergent spectrum and light-curve shape depend sensitively on radiative transport through the expanding outflow. Addressing how QPE flares are produced will require radiation-hydrodynamics simulations of stream-disk collisions, analogous to recent works by \citet{Huang2025} and \citet{Vurm2025} but in the hydrodynamic context advocated for here.

\subsection{One versus two flares per orbit}

A puzzle in QPE phenomenology is whether the observed recurrence time corresponds to a full orbital period of an interacting secondary, or to half an orbital period because two flares occur per orbit. This question is also examined analytically in Section~7.2 of \citet{Linial2025};
here we revisit it in light of our simulations. Observationally, a subset of QPE sources exhibit a clear alternating long-short recurrence pattern, sometimes correlated with strong-weak burst pairs. In the simplest geometric interpretation, such alternation is naturally produced when the emitter is tied to an inclined, mildly eccentric orbit that intersects the disk twice per orbit at the ascending and descending nodes: the time between successive node passages is generally unequal when $e\neq 0$, so the observed waiting times alternate, while the orbital period is given by $P_{\rm orb}\simeq \Delta t_{\rm long}+\Delta t_{\rm short}$. 

Within the star-debris-disk picture explored here, the stream-disk interaction provides a clean separation between (i) the clock (set by the orbit) and (ii) the flare duration (set by the stream crossing time through the disk). On the surface level, this makes both one- and two-flare-per-orbit interpretations viable without altering the key result that the duty cycle is $\sim$10--20\% once the flare duration tracks the stream length. However, diluted stellar debris streams are most likely significantly less dense than the disk and hence do not penetrate through it except for very small $a/\rt$ (also discussed in \S \ref{sec:duration}). For most viewing angles (except for observing the system edge-on), the wind-like emission from shocked debris is likely blocked by the extended optically thick disk and so only observable once per orbit. This expectation has recently received independent empirical support: using $O\!-\!C$ timing analysis of the regular source eRO-QPE2, \citet{Arcodia2026} find that the correlated odd/even eruption delays disfavor scenarios in which both disk crossings per orbit are observed, concluding that EMRI--disk collision models are viable only if a single event per orbit is observed. Hence, if most QPE systems are powered by debris-disk interactions, there is likely only one observable flare from the shocked stellar debris per orbit.

One possibility for producing two flares per orbit independent of viewing angle is for the shocked disk and stellar debris mass to be comparable so that one observed flare is from the shocked stellar debris and one from the shocked disk.   We do find that this is increasingly likely to be true at the smallest $a/r_{\rm t}$, i.e., shortest orbital period: in ART3.5 the shocked and unbound disk mass becomes comparable to (and in the unbound case exceeds) the stellar debris component (Table~\ref{tab:masses}), whereas at larger $a/r_{\rm t}$ the stellar debris dominates. This could explain the prominence of the long-short pattern in the short recurrence time sources GSN 069 and eRO-QPE2. However, naively this proposal is fine-tuned because in general the shocked disk and shocked debris have quite different physical conditions, and it seems unlikely that the two quite different shocked components would have nearly identical radiative output (as required in a two-flare-per-orbit scenario given that adjacent flares have quite similar luminosities and durations). It is also in tension with \citet{Arcodia2026}, whose timing analysis of eRO-QPE2 -- the shortest-period known QPE, and thus the most promising candidate for comparable shocked masses -- instead favors a single observed event per orbit. We defer this to future work in radiation hydrodynamics simulations.  We do note that in principle our model predicts that all QPEs should have two flares of {\em some kind} per orbit, even at longer orbital periods, one powered by the shocked stellar debris and one powered by the shocked disk (see Fig. \ref{fig:schematic}).   Naively, we again would expect the shocked disk flare to be much fainter and shorter than the shocked stellar debris flare and it is not clear if both are equally efficient at producing X-rays.   Stacking of QPE light curves at longer period could reveal this faint emission component `between' flares that is produced by the shocked disk debris.   

Beyond the geometric timing asymmetry noted at the start of this section, even a small orbital eccentricity could imprint a long-short pattern through differences in the debris \textit{stream} produced at each crossing, an effect absent from our circular-orbit simulations. For $e \neq 0$, the two disk crossings per orbit occur at different orbital phases and distances from the SMBH, so the collision velocity $v_\star$ and local disk surface density $\Sigma_{\rm disk}$ differ between the two nodes (cf. Eq.~\ref{eq:mstrip}). Because the stream's geometry and asymmetry are set by the ejecta velocity of the stripped material (\S\ref{sec:results}; Fig.~\ref{fig:stream}), these phase-dependent collision conditions produce two streams with different stripped masses, densities, and, in particular, different centers of mass and elongations relative to the Hill sphere. The stream formed at one crossing therefore presents a different effective cross section and column to the disk when it returns to collide half an orbit later than the stream formed at the other crossing. This offers an additional avenue for an apparent long-short alternation in flare luminosity and duration, distinct from (and superimposed on) the timing alternation that arises purely from the unequal $\Delta t_{\rm long}$ and $\Delta t_{\rm short}$ between pericenter and apocenter node passages. Disentangling these two contributions will require eccentric-orbit simulations that self-consistently track the stream properties produced at each node, which we defer to future work.

If most systems show only one observable flare per orbit, the debris-disk collision picture must still account for the long-short alternation seen in a subset of QPEs without requiring two observable node-crossing events. Beyond the eccentricity-driven stream differences described above, precession, realistic (e.g.\ non-axisymmetric or warped) disk structure, and viewing-angle--dependent radiative transfer could further break the symmetry between the two crossings and yield an alternating cadence. Our idealized circular-orbit setup captures none of these effects; testing them will require more global, eccentric-orbit simulations.

\subsection{Connections to other repeating nuclear transients: ASASSN-14ko and AT2023uqm}
QPEs occupy the short-period end of a broader landscape of repeating nuclear transients that plausibly share a common ingredient: a star on a bound orbit undergoing repeated strong interactions with the SMBH environment (partial stripping, mass transfer, or disk crossing). ASASSN-14ko is the prototypical long-period (114-day) periodic nuclear transient with UV/optical flares and evidence for period evolution \citep{Payne2021}; it is widely interpreted in terms of repeated partial tidal disruption (or closely related periodic mass-loss mechanisms). The observed high period derivative seems to necessitate disk drag \citep{Linial2024}. AT2023uqm extends this class by exhibiting multiple periodic optical flares with a ``runaway” increase in energy and a repeated double-peaked flare structure, suggestive of progressively strengthening partial disruptions and/or two interaction episodes per orbit \citep{Wang2025}.

Although the radiative bandpass differs (optical/UV for ASASSN-14ko and AT2023uqm versus soft X-rays for QPEs), the underlying dynamics may be continuous across parameter space. In particular, changing the SMBH mass, pericenter distance, and the availability/structure of an inner gaseous disk can shift (i) the orbital period from days to months, (ii) the dominant dissipation channel from stream circularization and reprocessing (optical/UV) to collision-powered, photon-starved shocks (soft X-ray), and (iii) the duty cycle and flare morphology. Recent discoveries of repeating X-ray transients on intermediate ($\sim$days) timescales suggest that such a continuum may already be emerging observationally. In this broader view, QPEs represent the regime where (a) the interaction repeats on short periods and (b) a compact, optically thick, inner flow/disk is present to efficiently shock, trap, and reprocess energy into the soft X-ray band.

\section*{Acknowledgments}
We thank Riccardo Arcodia, Joheen Chakraborty, Xiaoshan Huang, Wenbin Lu, Andy Mummery, and Brian Metzger for useful conversations. This work was supported in part by the Gordon and Betty Moore Foundation through grant GBMF5076. It benefited from interactions at the Kavli Institute for Theoretical Physics, supported by NSF PHY-2309135. Support for this work was provided by NASA through the NASA Hubble Fellowship grant \#HST-HF2-51581.001-A awarded by the Space Telescope Science Institute, which is operated by the Association of Universities for Research in Astronomy, Inc., for NASA, under contract NAS5-26555.

\bibliography{sample631}{}

@ARTICLE{Linial2023a,
       author = {{Linial}, Itai and {Sari}, Re'em},
        title = "{Unstable Mass Transfer from a Main-sequence Star to a Supermassive Black Hole and Quasiperiodic Eruptions}",
      journal = {\apj},
         year = 2023,
        month = mar,
       volume = {945},
       number = {2},
          eid = {86},
        pages = {86},
          doi = {10.3847/1538-4357/acbd3d},
archivePrefix = {arXiv},
       eprint = {2211.09851},
 primaryClass = {astro-ph.HE},
       adsurl = {https://ui.adsabs.harvard.edu/abs/2023ApJ...945...86L}
}

@ARTICLE{Linial2024b,
       author = {{Linial}, Itai and {Metzger}, Brian D.},
        title = "{Coupled Disk-star Evolution in Galactic Nuclei and the Lifetimes of QPE Sources}",
      journal = {\apj},
         year = 2024,
        month = oct,
       volume = {973},
       number = {2},
          eid = {101},
        pages = {101},
          doi = {10.3847/1538-4357/ad639e},
archivePrefix = {arXiv},
       eprint = {2404.12421},
 primaryClass = {astro-ph.HE},
       adsurl = {https://ui.adsabs.harvard.edu/abs/2024ApJ...973..101L}
}

@ARTICLE{Linial2026,
       author = {{Linial}, Itai and {Metzger}, Brian D. and {Beloborodov}, Andrei M.},
        title = "{Delayed Radio Flares in Tidal Disruption Events from Star-Disk Collision Outflows}",
      journal = {arXiv e-prints},
         year = 2026,
        month = may,
          eid = {arXiv:2605.28928},
        pages = {arXiv:2605.28928},
          doi = {10.48550/arXiv.2605.28928},
archivePrefix = {arXiv},
       eprint = {2605.28928},
 primaryClass = {astro-ph.HE},
       adsurl = {https://ui.adsabs.harvard.edu/abs/2026arXiv260528928L}
}

@ARTICLE{Guo2026a,
       author = {{Guo}, Wenyuan and {Shen}, Rong-Feng},
        title = "{Testing the Shock-cooling Emission Model from Star─Disk Collisions for Quasiperiodic Eruptions}",
      journal = {\apj},
         year = 2026,
        month = feb,
       volume = {998},
       number = {1},
          eid = {78},
        pages = {78},
          doi = {10.3847/1538-4357/ae3562},
archivePrefix = {arXiv},
       eprint = {2504.12762},
 primaryClass = {astro-ph.HE},
       adsurl = {https://ui.adsabs.harvard.edu/abs/2026ApJ...998...78G}
}

@ARTICLE{Miniutti2019,
       author = {{Miniutti}, G. and {Saxton}, R.~D. and {Giustini}, M. and {Alexander}, K.~D. and {Fender}, R.~P. and {Heywood}, I. and {Monageng}, I. and {Coriat}, M. and {Tzioumis}, A.~K. and {Read}, A.~M. and {Knigge}, C. and {Gandhi}, P. and {Pretorius}, M.~L. and {Ag{\'\i}s-Gonz{\'a}lez}, B.},
        title = "{Nine-hour X-ray quasi-periodic eruptions from a low-mass black hole galactic nucleus}",
      journal = {\nat},
         year = 2019,
        month = sep,
       volume = {573},
       number = {7774},
        pages = {381-384},
          doi = {10.1038/s41586-019-1556-x},
archivePrefix = {arXiv},
       eprint = {1909.04693},
 primaryClass = {astro-ph.HE},
       adsurl = {https://ui.adsabs.harvard.edu/abs/2019Natur.573..381M}
}

@ARTICLE{Guo2026,
       author = {{Guo}, Hengxiao and {Yan}, Zhen and {Li}, Ya-Ping and {Chakraborty}, Joheen and {S{\'a}nchez-S{\'a}ez}, Paula and {Hern{\'a}ndez-Garc{\'\i}a}, Lorena and {Zhang}, Wenda and {Sun}, Jingbo and {Li}, Shuang-Liang and {Deng}, Hongping and {Zuo}, Wenwen and {Tagawa}, Hiromichi and {Pan}, Xin and {Zhang}, Minghao and {Ar{\'e}valo}, Patricia and {Lira}, Paulina and {Jin}, Chichuan and {Gu}, Minfeng},
        title = "{Evidence for a Delayed Ultraviolet Counterpart to X-Ray Quasiperiodic Eruptions in Ansky}",
      journal = {\apjl},
         year = 2026,
        month = apr,
       volume = {1000},
       number = {2},
          eid = {L57},
        pages = {L57},
          doi = {10.3847/2041-8213/ae524b},
archivePrefix = {arXiv},
       eprint = {2603.02517},
 primaryClass = {astro-ph.HE},
       adsurl = {https://ui.adsabs.harvard.edu/abs/2026ApJ..1000L..57G}
}

@ARTICLE{Payne2021,
       author = {{Payne}, Anna V. and {Shappee}, Benjamin J. and {Hinkle}, Jason T. and {Vallely}, Patrick J. and {Kochanek}, Christopher S. and {Holoien}, Thomas W. -S. and {Auchettl}, Katie and {Stanek}, K.~Z. and {Thompson}, Todd A. and {Neustadt}, Jack M.~M. and {Tucker}, Michael A. and {Armstrong}, James D. and {Brimacombe}, Joseph and {Cacella}, Paulo and {Cornect}, Robert and {Denneau}, Larry and {Fausnaugh}, Michael M. and {Flewelling}, Heather and {Grupe}, Dirk and {Heinze}, A.~N. and {Lopez}, Laura A. and {Monard}, Berto and {Prieto}, Jose L. and {Schneider}, Adam C. and {Sheppard}, Scott S. and {Tonry}, John L. and {Weiland}, Henry},
        title = "{ASASSN-14ko is a Periodic Nuclear Transient in ESO 253-G003}",
      journal = {\apj},
         year = 2021,
        month = apr,
       volume = {910},
       number = {2},
          eid = {125},
        pages = {125},
          doi = {10.3847/1538-4357/abe38d},
archivePrefix = {arXiv},
       eprint = {2009.03321},
 primaryClass = {astro-ph.HE},
       adsurl = {https://ui.adsabs.harvard.edu/abs/2021ApJ...910..125P}
}

@ARTICLE{Linial2024,
       author = {{Linial}, Itai and {Quataert}, Eliot},
        title = "{Period evolution of repeating transients in galactic nuclei}",
      journal = {\mnras},
         year = 2024,
        month = jan,
       volume = {527},
       number = {2},
        pages = {4317-4329},
          doi = {10.1093/mnras/stad3470},
archivePrefix = {arXiv},
       eprint = {2309.15849},
 primaryClass = {astro-ph.HE},
       adsurl = {https://ui.adsabs.harvard.edu/abs/2024MNRAS.527.4317L}
}

@ARTICLE{Lu2023,
       author = {{Lu}, Wenbin and {Quataert}, Eliot},
        title = "{Quasi-periodic eruptions from mildly eccentric unstable mass transfer in galactic nuclei}",
      journal = {\mnras},
         year = 2023,
        month = oct,
       volume = {524},
       number = {4},
        pages = {6247-6266},
          doi = {10.1093/mnras/stad2203},
archivePrefix = {arXiv},
       eprint = {2210.08023},
 primaryClass = {astro-ph.HE},
       adsurl = {https://ui.adsabs.harvard.edu/abs/2023MNRAS.524.6247L}
}

@ARTICLE{Linial2023,
       author = {{Linial}, Itai and {Metzger}, Brian D.},
        title = "{EMRI + TDE = QPE: Periodic X-Ray Flares from Star-Disk Collisions in Galactic Nuclei}",
      journal = {\apj},
         year = 2023,
        month = nov,
       volume = {957},
       number = {1},
          eid = {34},
        pages = {34},
          doi = {10.3847/1538-4357/acf65b},
archivePrefix = {arXiv},
       eprint = {2303.16231},
 primaryClass = {astro-ph.HE},
       adsurl = {https://ui.adsabs.harvard.edu/abs/2023ApJ...957...34L}
}

@ARTICLE{Jiang2022,
       author = {{Jiang}, Yan-Fei},
        title = "{Multigroup Radiation Magnetohydrodynamics Based on Discrete Ordinates including Compton Scattering}",
      journal = {\apjs},
         year = 2022,
        month = nov,
       volume = {263},
       number = {1},
          eid = {4},
        pages = {4},
          doi = {10.3847/1538-4365/ac9231},
archivePrefix = {arXiv},
       eprint = {2209.06240},
 primaryClass = {astro-ph.IM},
       adsurl = {https://ui.adsabs.harvard.edu/abs/2022ApJS..263....4J}
}

@ARTICLE{Krolik2022,
       author = {{Krolik}, Julian H. and {Linial}, Itai},
        title = "{Quasiperiodic Erupters: A Stellar Mass-transfer Model for the Radiation}",
      journal = {\apj},
         year = 2022,
        month = dec,
       volume = {941},
       number = {1},
          eid = {24},
        pages = {24},
          doi = {10.3847/1538-4357/ac9eb6},
archivePrefix = {arXiv},
       eprint = {2209.02786},
 primaryClass = {astro-ph.HE},
       adsurl = {https://ui.adsabs.harvard.edu/abs/2022ApJ...941...24K}
}

@ARTICLE{Arcodia2021,
       author = {{Arcodia}, R. and {Merloni}, A. and {Nandra}, K. and {Buchner}, J. and {Salvato}, M. and {Pasham}, D. and {Remillard}, R. and {Comparat}, J. and {Lamer}, G. and {Ponti}, G. and {Malyali}, A. and {Wolf}, J. and {Arzoumanian}, Z. and {Bogensberger}, D. and {Buckley}, D.~A.~H. and {Gendreau}, K. and {Gromadzki}, M. and {Kara}, E. and {Krumpe}, M. and {Markwardt}, C. and {Ramos-Ceja}, M.~E. and {Rau}, A. and {Schramm}, M. and {Schwope}, A.},
        title = "{X-ray quasi-periodic eruptions from two previously quiescent galaxies}",
      journal = {\nat},
         year = 2021,
        month = apr,
       volume = {592},
       number = {7856},
        pages = {704-707},
          doi = {10.1038/s41586-021-03394-6},
archivePrefix = {arXiv},
       eprint = {2104.13388},
 primaryClass = {astro-ph.HE},
       adsurl = {https://ui.adsabs.harvard.edu/abs/2021Natur.592..704A}
}

@ARTICLE{Wevers2022,
       author = {{Wevers}, T. and {Pasham}, D.~R. and {Jalan}, P. and {Rakshit}, S. and {Arcodia}, R.},
        title = "{Host galaxy properties of quasi-periodically erupting X-ray sources}",
      journal = {\aap},
         year = 2022,
        month = mar,
       volume = {659},
          eid = {L2},
        pages = {L2},
          doi = {10.1051/0004-6361/202243143},
archivePrefix = {arXiv},
       eprint = {2201.11751},
 primaryClass = {astro-ph.HE},
       adsurl = {https://ui.adsabs.harvard.edu/abs/2022A&A...659L...2W}
}

@ARTICLE{Giustini2020,
       author = {{Giustini}, Margherita and {Miniutti}, Giovanni and {Saxton}, Richard D.},
        title = "{X-ray quasi-periodic eruptions from the galactic nucleus of RX J1301.9+2747}",
      journal = {\aap},
         year = 2020,
        month = apr,
       volume = {636},
          eid = {L2},
        pages = {L2},
          doi = {10.1051/0004-6361/202037610},
archivePrefix = {arXiv},
       eprint = {2002.08967},
 primaryClass = {astro-ph.HE},
       adsurl = {https://ui.adsabs.harvard.edu/abs/2020A&A...636L...2G}
}

@ARTICLE{Chakraborty2021,
       author = {{Chakraborty}, Joheen and {Kara}, Erin and {Masterson}, Megan and {Giustini}, Margherita and {Miniutti}, Giovanni and {Saxton}, Richard},
        title = "{Possible X-Ray Quasi-periodic Eruptions in a Tidal Disruption Event Candidate}",
      journal = {\apjl},
         year = 2021,
        month = nov,
       volume = {921},
       number = {2},
          eid = {L40},
        pages = {L40},
          doi = {10.3847/2041-8213/ac313b},
archivePrefix = {arXiv},
       eprint = {2110.10786},
 primaryClass = {astro-ph.HE},
       adsurl = {https://ui.adsabs.harvard.edu/abs/2021ApJ...921L..40C}
}

@ARTICLE{Arcodia2024,
       author = {{Arcodia}, R. and {Liu}, Z. and {Merloni}, A. and {Malyali}, A. and {Rau}, A. and {Chakraborty}, J. and {Goodwin}, A. and {Buckley}, D. and {Brink}, J. and {Gromadzki}, M. and {Arzoumanian}, Z. and {Buchner}, J. and {Kara}, E. and {Nandra}, K. and {Ponti}, G. and {Salvato}, M. and {Anderson}, G. and {Baldini}, P. and {Grotova}, I. and {Krumpe}, M. and {Maitra}, C. and {Miller-Jones}, J.~C.~A. and {Ramos-Ceja}, M.~E.},
        title = "{The more the merrier: SRG/eROSITA discovers two further galaxies showing X-ray quasi-periodic eruptions}",
      journal = {\aap},
         year = 2024,
        month = apr,
       volume = {684},
          eid = {A64},
        pages = {A64},
          doi = {10.1051/0004-6361/202348881},
archivePrefix = {arXiv},
       eprint = {2401.17275},
 primaryClass = {astro-ph.HE},
       adsurl = {https://ui.adsabs.harvard.edu/abs/2024A&A...684A..64A}
}

@ARTICLE{Raj2021,
       author = {{Raj}, A. and {Nixon}, C.~J.},
        title = "{Disk Tearing: Implications for Black Hole Accretion and AGN Variability}",
      journal = {\apj},
         year = 2021,
        month = mar,
       volume = {909},
       number = {1},
          eid = {82},
        pages = {82},
          doi = {10.3847/1538-4357/abdc25},
archivePrefix = {arXiv},
       eprint = {2101.05825},
 primaryClass = {astro-ph.HE},
       adsurl = {https://ui.adsabs.harvard.edu/abs/2021ApJ...909...82R}
}

@ARTICLE{Kaur2023,
       author = {{Kaur}, Karamveer and {Stone}, Nicholas C. and {Gilbaum}, Shmuel},
        title = "{Magnetically dominated discs in tidal disruption events and quasi-periodic eruptions}",
      journal = {\mnras},
         year = 2023,
        month = sep,
       volume = {524},
       number = {1},
        pages = {1269-1290},
          doi = {10.1093/mnras/stad1894},
archivePrefix = {arXiv},
       eprint = {2211.00704},
 primaryClass = {astro-ph.HE},
       adsurl = {https://ui.adsabs.harvard.edu/abs/2023MNRAS.524.1269K}
}

@ARTICLE{Xian2021,
       author = {{Xian}, Jingtao and {Zhang}, Fupeng and {Dou}, Liming and {He}, Jiasheng and {Shu}, Xinwen},
        title = "{X-Ray Quasi-periodic Eruptions Driven by Star-Disk Collisions: Application to GSN069 and Probing the Spin of Massive Black Holes}",
      journal = {\apjl},
         year = 2021,
        month = nov,
       volume = {921},
       number = {2},
          eid = {L32},
        pages = {L32},
          doi = {10.3847/2041-8213/ac31aa},
archivePrefix = {arXiv},
       eprint = {2110.10855},
 primaryClass = {astro-ph.HE},
       adsurl = {https://ui.adsabs.harvard.edu/abs/2021ApJ...921L..32X}
}

@ARTICLE{Sukova2021,
       author = {{Sukov{\'a}}, Petra and {Zaja{\v{c}}ek}, Michal and {Witzany}, Vojt{\v{e}}ch and {Karas}, Vladim{\'\i}r},
        title = "{Stellar Transits across a Magnetized Accretion Torus as a Mechanism for Plasmoid Ejection}",
      journal = {\apj},
         year = 2021,
        month = aug,
       volume = {917},
       number = {1},
          eid = {43},
        pages = {43},
          doi = {10.3847/1538-4357/ac05c6},
archivePrefix = {arXiv},
       eprint = {2102.08135},
 primaryClass = {astro-ph.HE},
       adsurl = {https://ui.adsabs.harvard.edu/abs/2021ApJ...917...43S}
}

@ARTICLE{Tagawa2023,
       author = {{Tagawa}, Hiromichi and {Haiman}, Zolt{\'a}n},
        title = "{Flares from stars crossing active galactic nucleus discs on low-inclination orbits}",
      journal = {\mnras},
         year = 2023,
        month = nov,
       volume = {526},
       number = {1},
        pages = {69-79},
          doi = {10.1093/mnras/stad2616},
archivePrefix = {arXiv},
       eprint = {2304.03670},
 primaryClass = {astro-ph.HE},
       adsurl = {https://ui.adsabs.harvard.edu/abs/2023MNRAS.526...69T}
}

@ARTICLE{Franchini2023,
       author = {{Franchini}, Alessia and {Bonetti}, Matteo and {Lupi}, Alessandro and {Miniutti}, Giovanni and {Bortolas}, Elisa and {Giustini}, Margherita and {Dotti}, Massimo and {Sesana}, Alberto and {Arcodia}, Riccardo and {Ryu}, Taeho},
        title = "{Quasi-periodic eruptions from impacts between the secondary and a rigidly precessing accretion disc in an extreme mass-ratio inspiral system}",
      journal = {\aap},
         year = 2023,
        month = jul,
       volume = {675},
          eid = {A100},
        pages = {A100},
          doi = {10.1051/0004-6361/202346565},
archivePrefix = {arXiv},
       eprint = {2304.00775},
 primaryClass = {astro-ph.HE},
       adsurl = {https://ui.adsabs.harvard.edu/abs/2023A&A...675A.100F}
}

@ARTICLE{Zhao2022,
       author = {{Zhao}, Z.~Y. and {Wang}, Y.~Y. and {Zou}, Y.~C. and {Wang}, F.~Y. and {Dai}, Z.~G.},
        title = "{Quasi-periodic eruptions from the helium envelope of hydrogen-deficient stars stripped by supermassive black holes}",
      journal = {\aap},
         year = 2022,
        month = may,
       volume = {661},
          eid = {A55},
        pages = {A55},
          doi = {10.1051/0004-6361/202142519},
archivePrefix = {arXiv},
       eprint = {2109.03471},
 primaryClass = {astro-ph.HE},
       adsurl = {https://ui.adsabs.harvard.edu/abs/2022A&A...661A..55Z}
}

@ARTICLE{Stone2020,
       author = {{Stone}, James M. and {Tomida}, Kengo and {White}, Christopher J. and {Felker}, Kyle G.},
        title = "{The Athena++ Adaptive Mesh Refinement Framework: Design and Magnetohydrodynamic Solvers}",
      journal = {\apjs},
         year = 2020,
        month = jul,
       volume = {249},
       number = {1},
          eid = {4},
        pages = {4},
          doi = {10.3847/1538-4365/ab929b},
archivePrefix = {arXiv},
       eprint = {2005.06651},
 primaryClass = {astro-ph.IM},
       adsurl = {https://ui.adsabs.harvard.edu/abs/2020ApJS..249....4S}
}

@ARTICLE{Nicholl2024,
       author = {{Nicholl}, M. and {Pasham}, D.~R. and {Mummery}, A. and {Guolo}, M. and {Gendreau}, K. and {Dewangan}, G.~C. and {Ferrara}, E.~C. and {Remillard}, R. and {Bonnerot}, C. and {Chakraborty}, J. and {Hajela}, A. and {Dhillon}, V.~S. and {Gillan}, A.~F. and {Greenwood}, J. and {Huber}, M.~E. and {Janiuk}, A. and {Salvesen}, G. and {van Velzen}, S. and {Aamer}, A. and {Alexander}, K.~D. and {Angus}, C.~R. and {Arzoumanian}, Z. and {Auchettl}, K. and {Berger}, E. and {de Boer}, T. and {Cendes}, Y. and {Chambers}, K.~C. and {Chen}, T. -W. and {Chornock}, R. and {Fulton}, M.~D. and {Gao}, H. and {Gillanders}, J.~H. and {Gomez}, S. and {Gompertz}, B.~P. and {Fabian}, A.~C. and {Herman}, J. and {Ingram}, A. and {Kara}, E. and {Laskar}, T. and {Lawrence}, A. and {Lin}, C. -C. and {Lowe}, T.~B. and {Magnier}, E.~A. and {Margutti}, R. and {McGee}, S.~L. and {Minguez}, P. and {Moore}, T. and {Nathan}, E. and {Oates}, S.~R. and {Patra}, K.~C. and {Ramsden}, P. and {Ravi}, V. and {Ridley}, E.~J. and {Sheng}, X. and {Smartt}, S.~J. and {Smith}, K.~W. and {Srivastav}, S. and {Stein}, R. and {Stevance}, H.~F. and {Turner}, S.~G.~D. and {Wainscoat}, R.~J. and {Weston}, J. and {Wevers}, T. and {Young}, D.~R.},
        title = "{Quasi-periodic X-ray eruptions years after a nearby tidal disruption event}",
      journal = {arXiv e-prints},
         year = 2024,
        month = sep,
          eid = {arXiv:2409.02181},
        pages = {arXiv:2409.02181},
          doi = {10.48550/arXiv.2409.02181},
archivePrefix = {arXiv},
       eprint = {2409.02181},
 primaryClass = {astro-ph.HE},
       adsurl = {https://ui.adsabs.harvard.edu/abs/2024arXiv240902181N}
}

@ARTICLE{Yao2025,
       author = {{Yao}, Philippe Z. and {Quataert}, Eliot and {Jiang}, Yan-Fei and {Lu}, Wenbin and {White}, Christopher J.},
        title = "{Star‑Disk Collisions: Implications for Quasi-periodic Eruptions and Other Transients near Supermassive Black Holes}",
      journal = {\apj},
         year = 2025,
        month = jan,
       volume = {978},
       number = {1},
          eid = {91},
        pages = {91},
          doi = {10.3847/1538-4357/ad8911},
archivePrefix = {arXiv},
       eprint = {2407.14578},
 primaryClass = {astro-ph.HE},
       adsurl = {https://ui.adsabs.harvard.edu/abs/2025ApJ...978...91Y}
}

@ARTICLE{Linial2025,
       author = {{Linial}, Itai and {Metzger}, Brian D. and {Quataert}, Eliot},
        title = "{QPEs from EMRI Debris Streams Impacting Accretion Disks in Galactic Nuclei}",
      journal = {\apj},
         year = 2025,
        month = oct,
       volume = {991},
       number = {2},
          eid = {147},
        pages = {147},
          doi = {10.3847/1538-4357/adfa0e},
archivePrefix = {arXiv},
       eprint = {2506.10096},
 primaryClass = {astro-ph.HE},
       adsurl = {https://ui.adsabs.harvard.edu/abs/2025ApJ...991..147L}
}

@ARTICLE{Huang2025,
       author = {{Huang}, Xiaoshan and {Linial}, Itai and {Jiang}, Yan-Fei},
        title = "{Multi-band Emission from Star-Disk Collision and Implications for Quasi-Periodic Eruptions}",
      journal = {arXiv e-prints},
         year = 2025,
        month = jun,
          eid = {arXiv:2506.11231},
        pages = {arXiv:2506.11231},
          doi = {10.48550/arXiv.2506.11231},
archivePrefix = {arXiv},
       eprint = {2506.11231},
 primaryClass = {astro-ph.HE},
       adsurl = {https://ui.adsabs.harvard.edu/abs/2025arXiv250611231H}
}

@ARTICLE{Kosec2025,
       author = {{Kosec}, P. and {Kara}, E. and {Brenneman}, L. and {Chakraborty}, J. and {Giustini}, M. and {Miniutti}, G. and {Pinto}, C. and {Rogantini}, D. and {Arcodia}, R. and {Middleton}, M. and {Sacchi}, A.},
        title = "{Detection of a Highly Ionized Outflow in the Quasiperiodically Erupting Source GSN 069}",
      journal = {\apj},
         year = 2025,
        month = jan,
       volume = {978},
       number = {1},
          eid = {10},
        pages = {10},
          doi = {10.3847/1538-4357/ad9249},
archivePrefix = {arXiv},
       eprint = {2406.17105},
 primaryClass = {astro-ph.HE},
       adsurl = {https://ui.adsabs.harvard.edu/abs/2025ApJ...978...10K}
}

@ARTICLE{Arcodia2025,
       author = {{Arcodia}, R. and {Baldini}, P. and {Merloni}, A. and {Rau}, A. and {Nandra}, K. and {Chakraborty}, J. and {Goodwin}, A.~J. and {Page}, M.~J. and {Buchner}, J. and {Masterson}, M. and {Monageng}, I. and {Arzoumanian}, Z. and {Buckley}, D. and {Kara}, E. and {Ponti}, G. and {Ramos-Ceja}, M.~E. and {Salvato}, M. and {Gendreau}, K. and {Grotova}, I. and {Krumpe}, M.},
        title = "{SRG/eROSITA No. 5: Discovery of Quasiperiodic Eruptions Every {\ensuremath{\sim}}3.7 days from a Galaxy at z > 0.1}",
      journal = {\apj},
         year = 2025,
        month = aug,
       volume = {989},
       number = {1},
          eid = {13},
        pages = {13},
          doi = {10.3847/1538-4357/adec9b},
archivePrefix = {arXiv},
       eprint = {2506.17138},
 primaryClass = {astro-ph.HE},
       adsurl = {https://ui.adsabs.harvard.edu/abs/2025ApJ...989...13A}
}

@ARTICLE{Quintin2023,
       author = {{Quintin}, E. and {Webb}, N.~A. and {Guillot}, S. and {Miniutti}, G. and {Kammoun}, E.~S. and {Giustini}, M. and {Arcodia}, R. and {Soucail}, G. and {Clerc}, N. and {Amato}, R. and {Markwardt}, C.~B.},
        title = "{Tormund's return: Hints of quasi-periodic eruption features from a recent optical tidal disruption event}",
      journal = {\aap},
         year = 2023,
        month = jul,
       volume = {675},
          eid = {A152},
        pages = {A152},
          doi = {10.1051/0004-6361/202346440},
archivePrefix = {arXiv},
       eprint = {2306.00438},
 primaryClass = {astro-ph.HE},
       adsurl = {https://ui.adsabs.harvard.edu/abs/2023A&A...675A.152Q}
}

@ARTICLE{Bykov2025,
       author = {{Bykov}, S.~D. and {Gilfanov}, M.~R. and {Sunyaev}, R.~A. and {Medvedev}, P.~S.},
        title = "{Further evidence of quasi-periodic eruptions in a tidal disruption event AT2019vcb by SRG/eROSITA}",
      journal = {\mnras},
         year = 2025,
        month = jun,
       volume = {540},
       number = {1},
        pages = {30-36},
          doi = {10.1093/mnras/staf686},
archivePrefix = {arXiv},
       eprint = {2409.16908},
 primaryClass = {astro-ph.HE},
       adsurl = {https://ui.adsabs.harvard.edu/abs/2025MNRAS.540...30B}
}

@ARTICLE{HernandezGarcia2025,
       author = {{Hern{\'a}ndez-Garc{\'\i}a}, Lorena and {Chakraborty}, Joheen and {S{\'a}nchez-S{\'a}ez}, Paula and {Ricci}, Claudio and {Cuadra}, Jorge and {McKernan}, Barry and {Ford}, K.~E. Saavik and {Ar{\'e}valo}, Patricia and {Rau}, Arne and {Arcodia}, Riccardo and {Kara}, Erin and {Liu}, Zhu and {Merloni}, Andrea and {Bruni}, Gabriele and {Goodwin}, Adelle and {Arzoumanian}, Zaven and {Assef}, Roberto J. and {Baldini}, Pietro and {Bayo}, Amelia and {Bauer}, Franz E. and {Bernal}, Santiago and {Brightman}, Murray and {Calistro Rivera}, Gabriela and {Gendreau}, Keith and {Homan}, David and {Krumpe}, Mirko and {Lira}, Paulina and {Mart{\'\i}nez-Aldama}, Mary Loli and {Salvato}, Mara and {Sotomayor}, Bel{\'e}n},
        title = "{Discovery of extreme quasi-periodic eruptions in a newly accreting massive black hole}",
      journal = {Nature Astronomy},
         year = 2025,
        month = jun,
       volume = {9},
        pages = {895-906},
          doi = {10.1038/s41550-025-02523-9},
archivePrefix = {arXiv},
       eprint = {2504.07169},
 primaryClass = {astro-ph.HE},
       adsurl = {https://ui.adsabs.harvard.edu/abs/2025NatAs...9..895H}
}

@ARTICLE{Vurm2025,
       author = {{Vurm}, Indrek and {Linial}, Itai and {Metzger}, Brian D.},
        title = "{Radiation Transport Simulations of Quasiperiodic Eruptions from Star{\textendash}Disk Collisions}",
      journal = {\apj},
         year = 2025,
        month = apr,
       volume = {983},
       number = {1},
          eid = {40},
        pages = {40},
          doi = {10.3847/1538-4357/adb74d},
archivePrefix = {arXiv},
       eprint = {2410.05166},
 primaryClass = {astro-ph.HE},
       adsurl = {https://ui.adsabs.harvard.edu/abs/2025ApJ...983...40V}
}

@ARTICLE{Stone2016,
       author = {{Stone}, Nicholas C. and {Metzger}, Brian D.},
        title = "{Rates of stellar tidal disruption as probes of the supermassive black hole mass function}",
      journal = {\mnras},
         year = 2016,
        month = jan,
       volume = {455},
       number = {1},
        pages = {859-883},
          doi = {10.1093/mnras/stv2281},
archivePrefix = {arXiv},
       eprint = {1410.7772},
 primaryClass = {astro-ph.HE},
       adsurl = {https://ui.adsabs.harvard.edu/abs/2016MNRAS.455..859S}
}

@ARTICLE{Yao2023,
       author = {{Yao}, Yuhan and {Ravi}, Vikram and {Gezari}, Suvi and {van Velzen}, Sjoert and {Lu}, Wenbin and {Schulze}, Steve and {Somalwar}, Jean J. and {Kulkarni}, S.~R. and {Hammerstein}, Erica and {Nicholl}, Matt and {Graham}, Matthew J. and {Perley}, Daniel A. and {Cenko}, S. Bradley and {Stein}, Robert and {Ricarte}, Angelo and {Chadayammuri}, Urmila and {Quataert}, Eliot and {Bellm}, Eric C. and {Bloom}, Joshua S. and {Dekany}, Richard and {Drake}, Andrew J. and {Groom}, Steven L. and {Mahabal}, Ashish A. and {Prince}, Thomas A. and {Riddle}, Reed and {Rusholme}, Ben and {Sharma}, Yashvi and {Sollerman}, Jesper and {Yan}, Lin},
        title = "{Tidal Disruption Event Demographics with the Zwicky Transient Facility: Volumetric Rates, Luminosity Function, and Implications for the Local Black Hole Mass Function}",
      journal = {\apjl},
         year = 2023,
        month = sep,
       volume = {955},
       number = {1},
          eid = {L6},
        pages = {L6},
          doi = {10.3847/2041-8213/acf216},
archivePrefix = {arXiv},
       eprint = {2303.06523},
 primaryClass = {astro-ph.HE},
       adsurl = {https://ui.adsabs.harvard.edu/abs/2023ApJ...955L...6Y}
}

@ARTICLE{Wang2025,
       author = {{Wang}, Yibo and {Wang}, Tingui and {Huang}, Shifeng and {Zhu}, Jiazheng and {Jiang}, Ning and {Lu}, Wenbin and {Shen}, Rongfeng and {Zhong}, Shiyan and {Lai}, Dong and {Yang}, Yi and {Shu}, Xinwen and {Xia}, Tianyu and {Luo}, Di and {Lyu}, Jianwei and {Brink}, Thomas and {Filippenko}, Alex and {Zheng}, Weikang and {Cai}, Minxuan and {Xu}, Zelin and {Wu}, Mingxin and {Zhang}, Xiaer and {Wu}, Weiyu and {Fan}, Lulu and {Jiang}, Ji-an and {Kong}, Xu and {Li}, Bin and {Lin}, Feng and {Liang}, Ming and {Luo}, Wentao and {Tang}, Jinlong and {Wan}, Zhen and {Wang}, Hairen and {Wang}, Jian and {Xue}, Yongquan and {Yao}, Dazhi and {Zhang}, Hongfei and {Zhao}, Wen and {Zheng}, Xianzhong and {Zhu}, Qingfeng and {Zuo}, Yingxi},
        title = "{A Star's Death by a Thousand Cuts: The Runaway Periodic Eruptions of AT2023uqm}",
      journal = {arXiv e-prints},
         year = 2025,
        month = oct,
          eid = {arXiv:2510.26561},
        pages = {arXiv:2510.26561},
          doi = {10.48550/arXiv.2510.26561},
archivePrefix = {arXiv},
       eprint = {2510.26561},
 primaryClass = {astro-ph.HE},
       adsurl = {https://ui.adsabs.harvard.edu/abs/2025arXiv251026561W}
}

@ARTICLE{Stone2010,
       author = {{Stone}, James M. and {Gardiner}, Thomas A.},
        title = "{Implementation of the Shearing Box Approximation in Athena}",
      journal = {\apjs},
         year = 2010,
        month = jul,
       volume = {189},
       number = {1},
        pages = {142-155},
          doi = {10.1088/0067-0049/189/1/142},
archivePrefix = {arXiv},
       eprint = {1006.0139},
 primaryClass = {astro-ph.IM},
       adsurl = {https://ui.adsabs.harvard.edu/abs/2010ApJS..189..142S}
}

@ARTICLE{Mummery2025,
       author = {{Mummery}, Andrew},
        title = "{Collisions with tidal disruption event disks: implications for quasi-periodic X-ray eruptions}",
      journal = {arXiv e-prints},
         year = 2025,
        month = apr,
          eid = {arXiv:2504.21456},
        pages = {arXiv:2504.21456},
          doi = {10.48550/arXiv.2504.21456},
archivePrefix = {arXiv},
       eprint = {2504.21456},
 primaryClass = {astro-ph.HE},
       adsurl = {https://ui.adsabs.harvard.edu/abs/2025arXiv250421456M}
}

@ARTICLE{Arcodia2026,
       author = {{Arcodia}, R. and {Miniutti}, G. and {Chakraborty}, J. and {Franchini}, A. and {Giustini}, M. and {Linial}, I. and {Mummery}, A. and {Bertassi}, L. and {Bonetti}, M. and {Kara}, E. and {Merloni}, A. and {Motta}, A. and {Ponti}, G. and {Quintin}, E. and {Soria}, R. and {Baldini}, P. and {Buchner}, J. and {Dotti}, M. and {Fragile}, P.~C. and {Ingram}, A. and {Middleton}, M. and {Panagiotou}, C. and {Sesana}, A. and {Yao}, P. and {Rau}, A. and {Vincentelli}, F.~M. and {Guolo}, M. and {Saxton}, R.},
        title = "{Even a Precessing Clock Is Right Twice per Orbit{\textemdash}The Superperiods of eRO-QPE2 and Challenges for Quasiperiodic Eruption Orbital Models}",
      journal = {\apj},
         year = 2026,
        month = jun,
       volume = {1003},
       number = {2},
          eid = {148},
        pages = {148},
          doi = {10.3847/1538-4357/ae6078},
archivePrefix = {arXiv},
       eprint = {2604.09788},
 primaryClass = {astro-ph.HE},
       adsurl = {https://ui.adsabs.harvard.edu/abs/2026ApJ..1003..148A}
}

@ARTICLE{Chakraborty2025,
       author = {{Chakraborty}, Joheen and {Kara}, Erin and {Arcodia}, Riccardo and {Buchner}, Johannes and {Giustini}, Margherita and {Hern{\'a}ndez-Garc{\'\i}a}, Lorena and {Linial}, Itai and {Masterson}, Megan and {Miniutti}, Giovanni and {Mummery}, Andrew and {Panagiotou}, Christos and {Quintin}, Erwan and {S{\'a}nchez-S{\'a}ez}, Paula},
        title = "{Discovery of Quasiperiodic Eruptions in the Tidal Disruption Event and Extreme Coronal Line Emitter AT2022upj: Implications for the QPE/TDE Fraction and a Connection to ECLEs}",
      journal = {\apjl},
         year = 2025,
        month = apr,
       volume = {983},
       number = {2},
          eid = {L39},
        pages = {L39},
          doi = {10.3847/2041-8213/adc2f8},
archivePrefix = {arXiv},
       eprint = {2503.19013},
 primaryClass = {astro-ph.HE},
       adsurl = {https://ui.adsabs.harvard.edu/abs/2025ApJ...983L..39C}
}

@ARTICLE{Baldini2026,
       author = {{Baldini}, P. and {Rau}, A. and {Merloni}, A. and {Trakhtenbrot}, B. and {Arcodia}, R. and {Giustini}, M. and {Miniutti}, G. and {Brennan}, S.~J. and {Freyberg}, M. and {S{\'a}nchez-S{\'a}ez}, P. and {Grotova}, I. and {Liu}, Z. and {Lian}, T. and {Nandra}, K.},
        title = "{Discovery of crested quasi-periodic eruptions following the most luminous SRG/eROSITA tidal disruption event}",
      journal = {\aap},
         year = 2026,
        month = feb,
       volume = {706},
          eid = {L15},
        pages = {L15},
          doi = {10.1051/0004-6361/202558241},
archivePrefix = {arXiv},
       eprint = {2602.03932},
 primaryClass = {astro-ph.HE},
       adsurl = {https://ui.adsabs.harvard.edu/abs/2026A&A...706L..15B}
}
\bibliographystyle{aasjournal}

\end{CJK*}

\end{document}